\documentclass[manuscript]{aastex63} 

\received{  }
\revised{  }
\accepted{   }
\submitjournal{ApJ}

\shorttitle{FAUST IV}
\shortauthors{Imai et al.}

\usepackage{longtable}
\usepackage{lineno}

\usepackage{rotating}

\newlength\maximageheight
\newlength\maximagewidth
\newlength\currentimagewidth
\newsavebox\imagebox

\newlength\maxsidewaysheight
\newcommand{\includesidewaysimg}[2][]{%
 \setlength\maximagewidth{\textheight}%
 \setlength\maximageheight{\maxsidewaysheight}%
 \innerincludeimg{#1}{#2}%
}

\newcommand{\innerincludeimg}[2]{%
 \sbox\imagebox{\includegraphics[#1,height=\maximageheight]{#2}}%
 \settowidth{\currentimagewidth}{\usebox\imagebox}%
 \ifdim\currentimagewidth>\maximagewidth
  \includegraphics[#1,width=\maximagewidth]{#2}%
 \else
  \usebox\imagebox
 \fi
}

\usepackage{graphicx}
\usepackage{here}
\usepackage{amssymb}

\newcommand{\dirfig}{figures_CB68/}
\renewcommand{\dirfig}{}

\newcommand{\bff}{}
\newcommand{\textbff}{}

\begin{document}

\title{Chemical and Physical Characterization of the Isolated Protostellar Source CB68: FAUST. IV}

\correspondingauthor{Yoko Oya}
\email{oya@taurus.phys.s.u-tokyo.ac.jp}





\author{Muneaki Imai}
\affiliation{Department of Physics, The University of Tokyo, 7-3-1, Hongo, Bunkyo-ku, Tokyo 113-0033, Japan}

\author{Yoko Oya}
\affiliation{Department of Physics, The University of Tokyo, 7-3-1, Hongo, Bunkyo-ku, Tokyo 113-0033, Japan}
\affiliation{Research Center for the Early Universe, The University of Tokyo, 7-3-1, Hongo, Bunkyo-ku, Tokyo 113-0033, Japan}

\author{Brian Svoboda}
\affiliation{National Radio Astronomy Observatory, PO Box O, Socorro, NM 87801, USA}
\affiliation{Jansky Fellow of the National Radio Astronomy Observatory}

\author{Hauyu Baobab Liu}
\affiliation{Institute of Astronomy and Astrophysics, Academia Sinica, 11F of Astronomy-Mathematics Building, AS/NTU No.1, Sec. 4, Roosevelt Rd., Taipei 10617, Taiwan, R.O.C.}

\author{Bertrand Lefloch}
\affiliation{Univ. Grenoble Alpes, CNRS, IPAG, 38000 Grenoble, France}

\author{Serena Viti}
\affiliation{Leiden Observatory, Leiden University, PO Box 9513, NL02300 RA,  Leiden, The Netherlands}

\author{Yichen Zhang}
\affiliation{RIKEN Cluster for Pioneering Research, 2-1, Hirosawa, Wako-shi, Saitama 351-0198, Japan}

\author{Cecilia Ceccarelli}
\affiliation{Univ. Grenoble Alpes, CNRS, IPAG, 38000 Grenoble, France}

\author{Claudio Codella}
\affiliation{INAF, Osservatorio Astrofisico di Arcetri, Largo E. Fermi 5, I-50125, Firenze, Italy}
\affiliation{Univ. Grenoble Alpes, CNRS, IPAG, 38000 Grenoble, France}

\author{Claire J. Chandler}
\affiliation{National Radio Astronomy Observatory, PO Box O, Socorro, NM 87801, USA}

\author{Nami Sakai}
\affiliation{RIKEN Cluster for Pioneering Research, 2-1, Hirosawa, Wako-shi, Saitama 351-0198, Japan}

\author{Yuri Aikawa}
\affiliation{Department of Astronomy, The University of Tokyo, 7-3-1 Hongo, Bunkyo-ku, Tokyo 113-0033, Japan}

\author{Felipe O. Alves}
\affiliation{Center for Astrochemical Studies, Max-Planck-Institut f\"{u}r extraterrestrische Physik (MPE), Gie$\beta$enbachstr. 1, D-85741 Garching, Germany}

\author{Nadia Balucani}
\affiliation{Department of Chemistry, Biology, and Biotechnology, The University of Perugia, Via Elce di Sotto 8, 06123 Perugia, Italy}

\author{Eleonora Bianchi}
\affiliation{Univ. Grenoble Alpes, CNRS, IPAG, 38000 Grenoble, France}

\author{Mathilde Bouvier}
\affiliation{Univ. Grenoble Alpes, CNRS, IPAG, 38000 Grenoble, France}

\author{Gemma Busquet}
\affiliation{Univ. Grenoble Alpes, CNRS, IPAG, 38000 Grenoble, France}
\affiliation{Departament de F\'{i}sica Qu\'{a}ntica i Astrof\'{i}sica, Institut de Ci\'{e}ncies del Cosmos, Universitat de Barcelona (IEEC-UB), Mart\'{i} i Franqu\'{e}s, 1, 08028, Barcelona, Catalunya, Spain}

\author{Paola Caselli}
\affiliation{Center for Astrochemical Studies, Max-Planck-Institut f\"{u}r extraterrestrische Physik (MPE), Gie$\beta$enbachstr. 1, D-85741 Garching, Germany}

\author{Emmanuel Caux}
\affiliation{IRAP, Universit\'{e} de Toulouse, CNRS, CNES, UPS, Toulouse, France}

\author{Steven Charnley}
\affiliation{Astrochemistry Laboratory, Code 691, NASA Goddard Space Flight Center, 8800 Greenbelt Road, Greenbelt, MD 20771, USA}

\author{Spandan Choudhury}
\affiliation{Center for Astrochemical Studies, Max-Planck-Institut f\"{u}r extraterrestrische Physik (MPE), Gie$\beta$enbachstr. 1, D-85741 Garching, Germany}

\author{Nicolas Cuello}
\affiliation{Univ. Grenoble Alpes, CNRS, IPAG, 38000 Grenoble, France}

\author{Marta De Simone}
\affiliation{Univ. Grenoble Alpes, CNRS, IPAG, 38000 Grenoble, France}

\author{Francois Dulieu}
\affiliation{CY Cergy Paris Universit\'{e}, Sorbonne Universit\'{e}, Observatoire de Paris, PSL University, CNRS, LERMA, F-95000, Cergy, France}

\author{Aurora Dur\'{a}n}
\affiliation{Instituto de Radioastronom\'{i}a y Astrof\'{i}sica , Universidad Nacional Aut\'{o}noma de M\'{e}xico, A.P. 3-72 (Xangari), 8701, Morelia, Mexico}

\author{Lucy Evans}
\affiliation{IRAP, Universit\'{e} de Toulouse, CNRS, CNES, UPS, Toulouse, France}
\affiliation{INAF, Osservatorio Astrofisico di Arcetri, Largo E. Fermi 5, I-50125, Firenze, Italy}

\author{C\'{e}cile Favre}
\affiliation{Univ. Grenoble Alpes, CNRS, IPAG, 38000 Grenoble, France}

\author{Davide Fedele}
\affiliation{INAF, Osservatorio Astrofisico di Arcetri, Largo E. Fermi 5, I-50125, Firenze, Italy}
\affiliation{INAF, Osservatorio Astrofisico di Torino, Via Osservatorio 20, 10025, Pino Torinese, Italy}

\author{Siyi Feng}
\affiliation{Department of Astronomy, Xiamen University, Xiamen, Fujian 361005, P. R. China}

\author{Francesco Fontani}
\affiliation{INAF, Osservatorio Astrofisico di Arcetri, Largo E. Fermi 5, I-50125, Firenze, Italy}
\affiliation{Center for Astrochemical Studies, Max-Planck-Institut f\"{u}r extraterrestrische Physik (MPE), Gie$\beta$enbachstr. 1, D-85741 Garching, Germany}

\author{Logan Francis}
\affiliation{NRC Herzberg Astronomy and Astrophysics, 5071 West Saanich Road, Victoria, BC, V9E 2E7, Canada}
\affiliation{Department of Physics and Astronomy, University of Victoria, Victoria, BC, V8P 5C2, Canada}

\author{Tetsuya Hama}
\affiliation{Komaba Institute for Science, The University of Tokyo, 3-8-1 Komaba, Meguro, Tokyo 153-8902, Japan}
\affiliation{Department of Basic Science, The University of Tokyo, 3-8-1 Komaba, Meguro, Tokyo 153-8902, Japan}

\author{Tomoyuki Hanawa}
\affiliation{Center for Frontier Science, Chiba University, 1-33 Yayoi-cho, Inage-ku, Chiba 263-8522, Japan}

\author{Eric Herbst}
\affiliation{Department of Chemistry, University of Virginia, McCormick Road, PO Box 400319, Charlottesville, VA 22904, USA}

\author{Shingo Hirano}
\affiliation{Department of Astronomy, The University of Tokyo, 7-3-1 Hongo, Bunkyo-ku, Tokyo 113-0033, Japan}

\author{Tomoya Hirota}
\affiliation{National Astronomical Observatory of Japan, Osawa 2-21-1, Mitaka-shi, Tokyo 181-8588, Japan}


\author{Andrea Isella}
\affiliation{Department of Physics and Astronomy, Rice University, 6100 Main Street, MS-108, Houston, TX 77005, USA}

\author{Izaskun J\'{i}menez-Serra}
\affiliation{Centro de Astrobiolog\'{\i}a (CSIC-INTA), Ctra. de Torrej\'on a Ajalvir, km 4, 28850, Torrej\'on de Ardoz, Spain}

\author{Doug Johnstone}
\affiliation{NRC Herzberg Astronomy and Astrophysics, 5071 West Saanich Road, Victoria, BC, V9E 2E7, Canada}
\affiliation{Department of Physics and Astronomy, University of Victoria, Victoria, BC, V8P 5C2, Canada}

\author{Claudine Kahane}
\affiliation{Univ. Grenoble Alpes, CNRS, IPAG, 38000 Grenoble, France}

\author{Romane Le Gal}
\affiliation{Univ. Grenoble Alpes, CNRS, IPAG, 38000 Grenoble, France}
\affiliation{Institut de Radioastronomie Millim\'{e}trique, 38406 Saint-Martin d'H$\grave{e}$res, France}

\author{Laurent Loinard}
\affiliation{Instituto de Radioastronom\'{i}a y Astrof\'{i}sica , Universidad Nacional Aut\'{o}noma de M\'{e}xico, A.P. 3-72 (Xangari), 8701, Morelia, Mexico}
\affiliation{Instituto de Astronom\'{i}a, Universidad Nacional Aut\'{o}noma de M\'{e}xico, Ciudad Universitaria, A.P. 70-264, Cuidad de M\'{e}xico 04510, Mexico}


\author{Ana L\'{o}pez-Sepulcre}
\affiliation{Univ. Grenoble Alpes, CNRS, IPAG, 38000 Grenoble, France}
\affiliation{Institut de Radioastronomie Millim\'{e}trique, 38406 Saint-Martin d'H$\grave{e}$res, France}

\author{Luke T. Maud}
\affiliation{European Southern Observatory, Karl-Schwarzschild Str. 2, 85748 Garching bei M\"{u}nchen, Germany}

\author{Mar\'{i}a Jos\'{e} Maureira}
\affiliation{Center for Astrochemical Studies, Max-Planck-Institut f\"{u}r extraterrestrische Physik (MPE), Gie$\beta$enbachstr. 1, D-85741 Garching, Germany}

\author{Francois Menard}
\affiliation{Univ. Grenoble Alpes, CNRS, IPAG, 38000 Grenoble, France}

\author{Seyma Mercimek}
\affiliation{INAF, Osservatorio Astrofisico di Arcetri, Largo E. Fermi 5, I-50125, Firenze, Italy}
\affiliation{Universit$\grave{a}$ degli Studi di Firenze, Dipartimento di Fisica e Astronomia, via G. Sansone 1, 50019 Sesto Fiorentino, Italy}

\author{Anna Miotello}
\affiliation{European Southern Observatory, Karl-Schwarzschild Str. 2, 85748 Garching bei M\"{u}nchen, Germany}

\author{George Moellenbrock}
\affiliation{National Radio Astronomy Observatory, PO Box O, Socorro, NM 87801, USA}

\author{Shoji Mori}
\affiliation{Department of Astronomy, The University of Tokyo, 7-3-1, Hongo, Bunkyo-ku, Tokyo 113-0033, Japan}

\author{Nadia M. Murillo}
\affiliation{RIKEN Cluster for Pioneering Research, 2-1, Hirosawa, Wako-shi, Saitama 351-0198, Japan}

\author{Riouhei Nakatani}
\affiliation{RIKEN Cluster for Pioneering Research, 2-1, Hirosawa, Wako-shi, Saitama 351-0198, Japan}

\author{Hideko Nomura}
\affiliation{Division of Science, National Astronomical Observatory of Japan, 2-21-1 Osawa, Mitaka, Tokyo 181-8588, Japan}

\author{Yasuhiro Oba}
\affiliation{Institute of Low Temperature Science, Hokkaido University, N19W8, Kita-ku, Sapporo, Hokkaido 060-0819, Japan}

\author{Ross O'Donoghue}
\affiliation{Department of Physics and Astronomy, University College London, Gower Street, London, WC1E 6BT, UK}

\author{Satoshi Ohashi}
\affiliation{RIKEN Cluster for Pioneering Research, 2-1, Hirosawa, Wako-shi, Saitama 351-0198, Japan}

\author{Yuki Okoda}
\affiliation{Department of Physics, The University of Tokyo, 7-3-1, Hongo, Bunkyo-ku, Tokyo 113-0033, Japan}
\affiliation{RIKEN Cluster for Pioneering Research, 2-1, Hirosawa, Wako-shi, Saitama 351-0198, Japan}

\author{Juan Ospina-Zamudio}
\affiliation{Univ. Grenoble Alpes, CNRS, IPAG, 38000 Grenoble, France}


\author{Jaime Pineda}
\affiliation{Center for Astrochemical Studies, Max-Planck-Institut f\"{u}r extraterrestrische Physik (MPE), Gie$\beta$enbachstr. 1, D-85741 Garching, Germany}

\author{Linda Podio}
\affiliation{INAF, Osservatorio Astrofisico di Arcetri, Largo E. Fermi 5, I-50125, Firenze, Italy}

\author{Albert Rimola}
\affiliation{Departament de Qu\'{i}mica, Universitat Aut$\grave{o}$noma de Barcelona, 08193 Bellaterra, Spain}

\author{Takeshi Sakai}
\affiliation{Graduate School of Informatics and Engineering, The University of Electro-Communications, Chofu, Tokyo 182-8585, Japan}

\author{Dominique Segura-Cox}
\affiliation{Center for Astrochemical Studies, Max-Planck-Institut f\"{u}r extraterrestrische Physik (MPE), Gie$\beta$enbachstr. 1, D-85741 Garching, Germany}

\author{Yancy Shirley}
\affiliation{Steward Observatory, 933 N Cherry Ave., Tucson, AZ 85721 USA}


\author{Vianney Taquet}
\affiliation{INAF, Osservatorio Astrofisico di Arcetri, Largo E. Fermi 5, I-50125, Firenze, Italy}

\author{Leonardo Testi}
\affiliation{European Southern Observatory, Karl-Schwarzschild Str. 2, 85748 Garching bei M\"{u}nchen, Germany}
\affiliation{INAF, Osservatorio Astrofisico di Arcetri, Largo E. Fermi 5, I-50125, Firenze, Italy}

\author{Charlotte Vastel}
\affiliation{IRAP, Universit\'{e} de Toulouse, CNRS, CNES, UPS, Toulouse, France}

\author{Naoki Watanabe}
\affiliation{Institute of Low Temperature Science, Hokkaido University, N19W8, Kita-ku, Sapporo, Hokkaido 060-0819, Japan}

\author{Yoshimasa Watanabe}
\affiliation{Materials Science and Engineering, College of Engineering, Shibaura Institute of Technology, 3-7-5 Toyosu, Koto-ku, Tokyo 135-8548, Japan}

\author{Arezu Witzel}
\affiliation{Univ. Grenoble Alpes, CNRS, IPAG, 38000 Grenoble, France}

\author{Ci Xue}
\affiliation{Department of Chemistry, University of Virginia, McCormick Road, PO Box 400319, Charlottesville, VA 22904, USA}
\affiliation{Department of Chemistry, Massachusetts Institute of Technology, 77 Massachusetts Avenue, Cambridge, MA 02139, USA}

\author{Bo Zhao}
\affiliation{Center for Astrochemical Studies, Max-Planck-Institut f\"{u}r extraterrestrische Physik (MPE), Gie$\beta$enbachstr. 1, D-85741 Garching, Germany}

\author{Satoshi Yamamoto}
\affiliation{Department of Physics, The University of Tokyo, 7-3-1, Hongo, Bunkyo-ku, Tokyo 113-0033, Japan}
\affiliation{Research Center for the Early Universe, The University of Tokyo, 7-3-1, Hongo, Bunkyo-ku, Tokyo 113-0033, Japan}

\begin{abstract}

Chemical diversity of low-mass protostellar sources has so far been recognized, and environmental effects are invoked as its origin. 
In this context, observations of isolated protostellar sources without influences of nearby objects are of particular importance. 
Here, we report chemical and physical structures of the low-mass Class 0 protostellar source IRAS 16544$-$1604 in the Bok globule CB68, based on 1.3 mm ALMA observations at a spatial resolution of $\sim$70~au that were conducted as part of the large program FAUST.
Three interstellar saturated complex organic molecules (iCOMs), CH$_3$OH, HCOOCH$_3$, and CH$_3$OCH$_3$, are detected toward the protostar. 
The rotation temperature and the emitting region size for CH$_3$OH are derived to be $131\pm11$~K and $\sim$10~au, respectively.
The detection of iCOMs in close proximity to the protostar indicates that CB68 harbors a hot corino. 
The kinematic structure of the C$^{18}$O, CH$_3$OH, and OCS lines is explained by an infalling-rotating envelope model, and the protostellar mass and the radius of the centrifugal barrier are estimated to be $0.08-0.30$~$M_\odot$ and $< 30$ au, respectively. 
The small radius of the centrifugal barrier seems to be related to the small emitting region of iCOMs.
In addition, we detect emission lines of c-C$_3$H$_2$ and CCH associated with the protostar, revealing a warm carbon chain chemistry (WCCC) on a 1000~au scale. 
We therefore find that the chemical structure of CB68 is described by a hybrid chemistry.
The molecular abundances are discussed in comparison with those in other hot corino sources and reported chemical models.

\end{abstract}


\section{Introduction} \label{sec:intro}
The chemical evolution from protostellar cores to protoplanetary disks remains an open problem of fundamental significance to understand the material origin of the solar system. 
In the last decade, significant diversity in the chemical composition of protostellar cores on a few 1000 au scale has been revealed \citep{sak13}. 
Two distinct cases are found: (1) hot corino chemistry, which is characterized by a richness of (saturated) interstellar complex organic molecules \citep[iCOMs;][]{caz03, her09, cas12}, and (2) warm carbon-chain chemistry (WCCC), which is characterized by a richness of (highly unsaturated) hydrocarbons such as carbon-chain molecules \citep{sak08}. 
The chemical diversity of protostellar disks has been investigated extensively, and it is now known that the diversity in initial chemical conditions seen on the core scales are inherited to disk scales 
\citep{sak14, oya16, oya18}. 
Such chemical diversity on core scales could originate from various environmental effects. The source location in a molecular cloud complex and association with nearby sources in particular can significantly affect the chemical composition 
\citep{lin15, spe16, hig18, oya19}. 
In order to understand the basic chemical evolution of protostellar sources, the investigation of isolated sources, not affected by nearby (proto-)stellar feedback, is of fundamental importance. Isolated protostellar sources are ideal laboratories not only for testing star formation theory
(e.g., \citealt{eva15}) but for testing theories of chemical evolution.

Recently, the chemical and physical characteristics of the representative isolated protostellar source B335 has been investigated with interferometric observations 
\citep{ima16, ima19}. 
They reported that B335 is characterized by a very compact distribution of iCOMs and a rather extended distribution of carbon-chain molecules. 
These properties are consistent with a hybrid chemistry composed of both hot corino chemistry and warm carbon-chain chemistry (WCCC) on different scales. 
Similar features have also been reported for L483 \citep{oya17}, 
and are consistent with the predictions of chemical network calculations of the gas-grain model \citep{aik08}. 
Moreover, the envelope gas in B335 is found to be in near free-fall with a very small rotation structure ($\sim 10$ au) recently identified \citep{ima19, bje19}. 
Now, it is important to examine whether these chemical and physical features found in B335 are also found in other isolated protostellar sources. 

We are conducting the ALMA large program FAUST (Fifty AU Study of the chemistry in the disk/envelope system of Solar-like protostars), 
exploring the physical and chemical structures on a 50 au scale toward 13 nearby young protostellar sources with the same brightness sensitivity \citep{cod21, bia20, oko21a, oha22}.
By observing more than 40 molecular species, FAUST systematically compares the physical and chemical structures of various types of sources. 
Observations of isolated sources are essential for this program.

CB68 (L146) is a small, nearby, slightly cometary-shaped opaque Bok globule located in the outskirts of the Ophiuchus molecular cloud complex 
\citep{noz91, lem96, lau97, lau98}. 
The canonical distance to the Ophiuchus molecular cloud complex of 160 pc has been used historically \citep{chi81, lau97, lau10}. 
However, \citet{lom08} reported the distance to the Ophiuchus molecular cloud complex to be $119 \pm 6$ pc by using the Hipparcos and Tycho parallax measurements with the 2MASS data. 
\citet{ort17} reported the distance of $137.3 \pm 1.2$ pc based on the VLBA maser parallax measurements and recommended the distance of 120$-$140 pc. 
Recently, \citet{zuc19} derived the distance to be $144 \pm 7$ pc based on extinction and parallax measurements by using Gaia data.  
Although controversy remains, we assume that CB68 belongs to the Ophiuchus molecular cloud complex, and use the distance of 140 pc in this paper. 

CB68 is a well-isolated protostellar source among the 32 sources in the catalog of Bok globules by 
\citet{lau10}. 
Since CB68 has a simple structure and can be accessed from ALMA, we selected this source as a target source of the FAUST program. 
The extent of the parent cloud core is about 0.03 pc, and its mass is estimated to be $\sim0.9$ $M_\odot$ \citep{ber14}. 
It harbors the Class 0 low-mass protostar IRAS 16544$-$1604, whose bolometric temperature and bolometric luminosity are 41 K and 0.86 $L_\odot$, respectively \citep{lau13}.  
The protostar IRAS 16544$-$1604 has been monitored in the mid-IR by NEOWISE \citep{mai14} over the last seven years. 
As part of a mid-IR  analysis of all known Gould Belt young stellar objects (YSOs), it was classified an irregular variable by \citet{par21} in that the observed variability across the 14 NEOWISE epochs is greater than three times the variability seen within an individual epoch. 
No NIR source is detected at 2.2 {$\mu$}m \citep{lau10}, indicating that no evolved protostars are associated with the globule. 
The protostar drives a collimated bipolar molecular outflow at P.A.\ 142\degr\ \citep{wuu96, val00}. 
In spite of these previous works, its detailed chemical and physical structures on a few 100 au scale or smaller are little understood. 
In this paper, we report the first chemical and physical characterization of this source on scales from 50 au to 1000 au. 

The paper is organized as follows.
After descriptions of observations in Section~\ref{sec:obs}, we present and discuss the chemical structure in Section~\ref{sec:chemstr}, where we compare the results with other low-mass protostellar sources and with gas-grain chemical models. Section~\ref{sec:kin} describes the kinematic structure around the protostar, and Section~\ref{sec:B335} the comparison of the chemical and physical features of CB68 with those of the other isolated source B335. Finally, Section~\ref{sec:summary} summarizes our main findings. 

\section{Observations} \label{sec:obs}
Observations of CB68 were conducted with ALMA (Cycle 6) in several sessions from 2018 October 4 to 2019 April 17 in the C43-1, C43-4, and ACA configurations, as part of the large program FAUST (2018.1.01205.L). 
The field center is $(\alpha_{\rm ICRS}, \delta_{\rm ICRS})$ = (16$^\mathrm{h}$57$^{\mathrm{m}}$19\fs643, $-$16\degr9\arcmin23\farcs92). 
Two frequency settings in Band 6, 216$-$234 GHz (Setup 1: 1.3~mm) and 245$-$260 GHz (Setup 2: 1.2~mm), were observed, 
where the ACA was employed to ensure a maximum recoverable size of $\sim 27$\arcsec\ ($\sim 3800$ au). The correlator was configured to observe 13 spectral windows.
The spectral window used for broad continuum coverage in the observations has a bandwidth of 1875 MHz and a channel width of 0.977 MHz.
This channel width corresponds to 
a velocity resolution of 
$\sim$1.25 km~s$^{-1}$ and $\sim$1.19 km~s$^{-1}$ for Setup 1 and Setup 2, respectively. 
The other twelve spectral windows are optimized for targeting specific molecular line transitions. These windows have a bandwidth of 58.6 MHz and spectral resolution of 141 kHz.
The channel width for the high resolution windows corresponds to a velocity resolution of $\sim 0.19$ km~s$^{-1}$ and $\sim 0.17$ km~s$^{-1}$ for Setup 1 and Setup 2, respectively. 

The Common Astronomy Software Applications (CASA) package \citep{mum07} was used for data reduction and imaging. The data were calibrated using a modified version of the ALMA calibration pipeline in CASA v5.6.1-8, with an additional in-house calibration routine to correct for the system temperature and spectral line data normalization\footnote{See \url{https://help.almascience.org/index.php?/Knowledgebase/Article/View/419}}. Common line-free frequencies across all configurations of a Setup were carefully selected and used to form a continuum dataset, which was then used for self-calibration. The phase self-calibration used a solution interval as short as possible to correct for tropospheric and systematic phase errors, at the same time as avoiding noisy solutions (typically 6 to 20 seconds). Long solution interval amplitude and phase self-calibration was also used to align amplitudes and positions across multiple datasets. The complex gain solutions were applied to the line visibilities, and the model of the continuum emission from the self-calibration was then subtracted from the line data, to produce self-calibrated, continuum-subtracted, line data. The uncertainty in the absolute flux density scale is approximately 10\%.

Maps of the spectral line emission were obtained by CLEANing the dirty images with the Briggs robustness parameter of 0.5.
The observation parameters of the line emission are summarized in Table~\ref{tab_CB68_COM_list}. 
Line parameters are taken from The Cologne Database for Molecular Spectroscopy (CDMS) \citep{end16} and the Jet Propulsion Laboratory (JPL) catalog \citep{pic98}.

\section{Chemical Structure} \label{sec:chemstr}

\subsection{Dust Continuum Emission} \label{sec:chemstr_dust}
\begin{figure}
	\centering
        \plotone{\dirfig 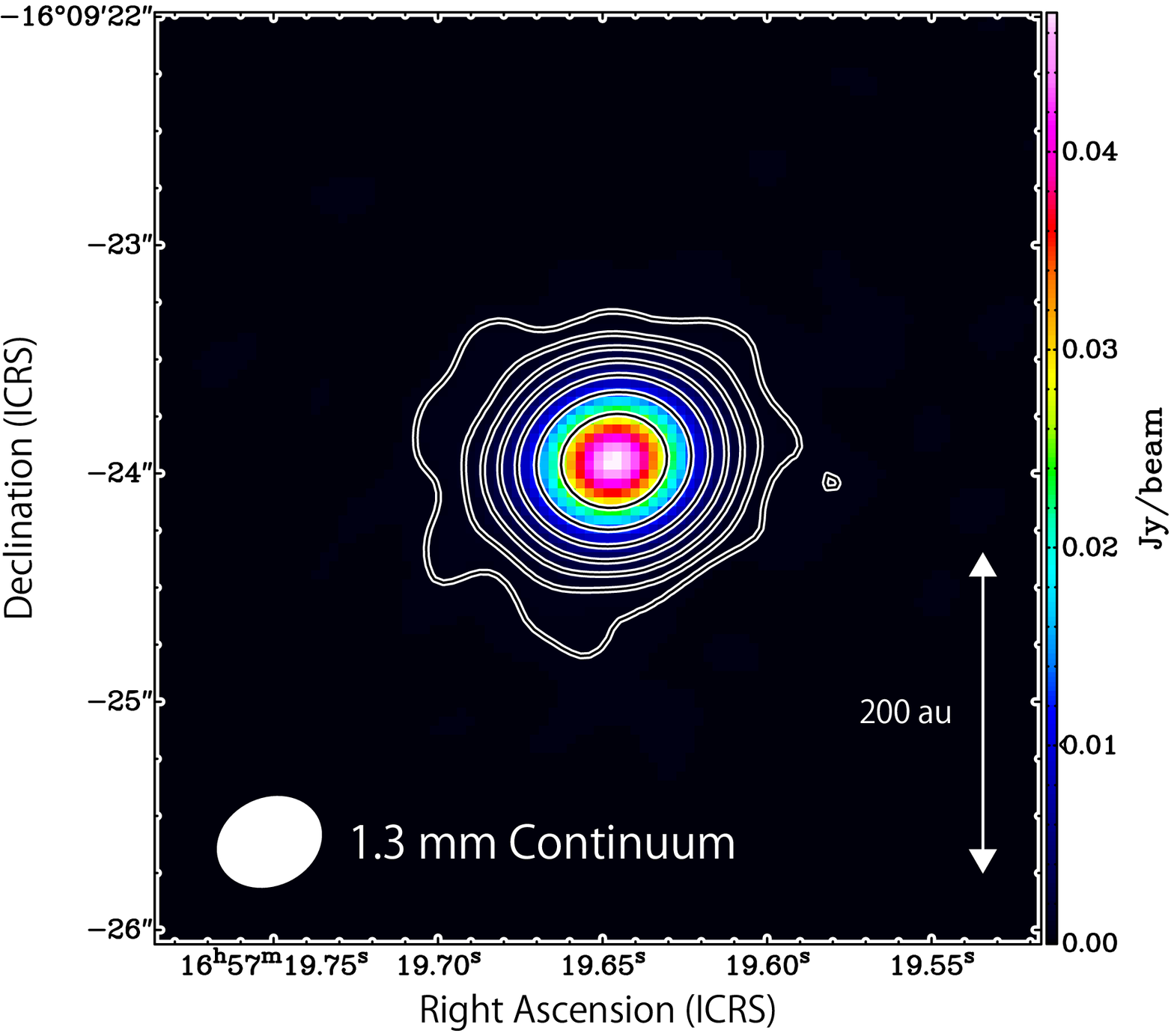}
	\caption{Images of the 1.3 mm dust continuum. 
			The peak position is ($\alpha_{IRCS}$, $\delta_{IRCS}$) = (16$^{\mathrm{h}}$57$^{\mathrm{m}}$19\fs647, $-$16\degr9\arcmin23\farcs94). The rms noise level ($\sigma$) is 40 $\mu$Jy beam$^{-1}$. The contours levels start at $10\sigma$ and double in value up to $640\sigma$. A white ellipse represents the beam size.}
        \label{fig:CB68_cont}
\end{figure}

Figure~\ref{fig:CB68_cont} shows the 1.3 mm continuum map whose beam size is 0\farcs47 $\times$ 0\farcs39 (P.A.: $-$65$.\!\!$\degr2). 
This has a higher angular resolution than the 1.2 mm continuum map, and hence, we here use the 1.3 mm data in the present study. 
The continuum emission is concentrated around the protostar. 
The peak position determined by a 2D Gaussian fit is ($\alpha_{\rm IRCS}$, $\delta_{\rm IRCS}$) =  (16$^{\mathrm{h}}$57$^{\mathrm{m}}$19\fs647, $-$16\degr09\arcmin23\farcs94). 
The integrated flux density and peak flux density are $55.9\pm1.1$ mJy and $47.1\pm0.6$ mJy beam$^{-1}$, respectively. 
The peak intensity corresponds to a brightness temperature of $5.73\pm0.07$ K. 
The source image size is (0\farcs473 $\pm$ 0\farcs006) $\times$ (0\farcs459 $\pm$ 0\farcs005) with P.A. of 123$^{\circ}$ $\pm$ 16$^{\circ}$. 
This corresponds the linear size of $(66.2 \pm 0.8)$ au $\times (64.2 \pm 0.8)$ au. Since the image size is comparable to the beam size, the deconvolution is not successful. The source is very compact. Detailed analyses of the dust emission will be described in a separate publication of FAUST. 

\subsection{Molecular Distributions} \label{sec:chemstr_mol-dist}

\begin{sidewaysfigure}
	\centering
	\includesidewaysimg{\dirfig 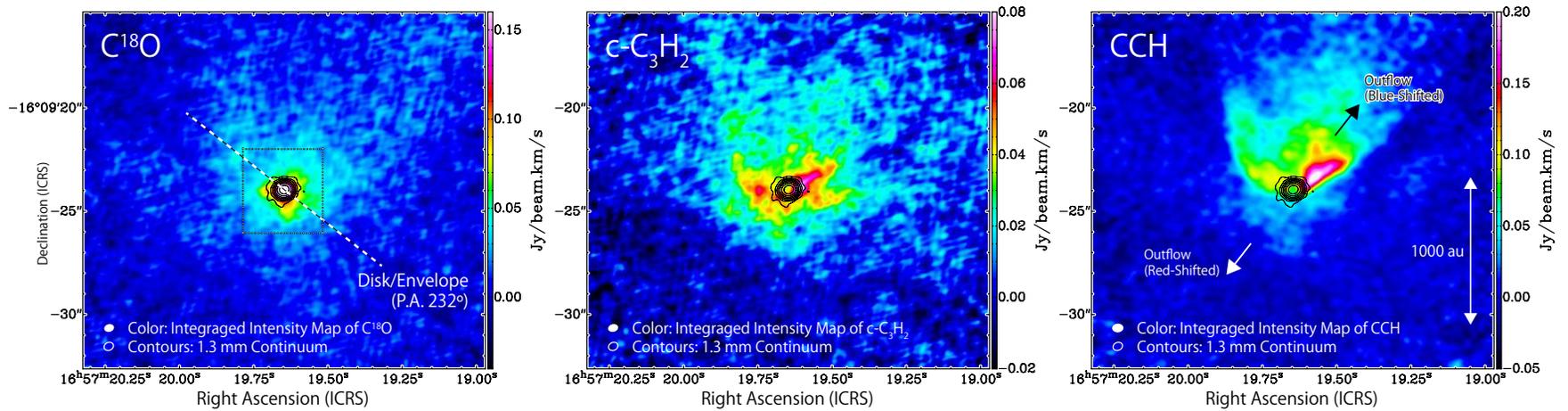} 
	\caption{Integrated intensity maps (moment 0 maps) of C$^{18}$O ($J = 2 - 1$), c-C$_3$H$_2$ ($6_{0,6} - 5_{1,5}$; $6_{1,6} - 5_{0,5}$), CCH ($N = 3 - 2, J = 7/2 - 5/2$) (Color) and the continuum image (black contours; Figure~\ref{fig:CB68_cont}). The integrated velocity range is from 0.0 km s$^{-1}$ to 9.8 km s$^{-1}$ for the three lines, where the systemic velocity is 5.0 km s$^{-1}$.  A dashed line on the left panel indicates the direction of the disk/envelope system which is perpendicular to the CO outflow direction (P.A. 142$^{\circ}$) \citep{wuu96, val00}. The rms noise level is 7 mJy beam$^{-1}$ km s$^{-1}$ for the three maps. A dashed rectangle in the C$^{18}$O panel indicates the area for Figures~\ref{fig:CB68_cont}, \ref{fig:CB68_moment0-COMs}, \ref{fig:CB68_mom1-PV}, and \ref{fig:CB68_OCS}. The beam sizes are shown on a bottom left corner of each panel.} 
	\label{fig:CB68_moment0-ext}
\end{sidewaysfigure}

Figure~\ref{fig:CB68_moment0-ext}a shows the integrated intensity map of the C$^{18}$O ($J=2-1$) line. The distribution of C$^{18}$O traces the protostellar core on a scale of 4\arcsec \ (560 au) around the protostar with its intensity peak coinciding the continuum position. Figures~\ref{fig:CB68_moment0-ext}b and \ref{fig:CB68_moment0-ext}c show the integrated intensity maps of the unsaturated hydrocarbon molecule lines, c-C$_3$H$_2$ ($6_{0,6}-5_{1,5}$, $6_{1,6}-5_{0,5}$) and CCH ($N=3-2$, $J=7/2-5/2$), respectively. The overall extent of the c-C$_3$H$_2$ emission is comparable to that of C$^{18}$O, indicating that c-C$_3$H$_2$ exists in the protostellar core traced by C$^{18}$O. Nevertheless, the distributions are different between c-C$_3$H$_2$ and C$^{18}$O. The c-C$_3$H$_2$ emission does not peak at the continuum peak position but shows a rather extended structure to the northwestern and eastern directions. Moreover, the weak emission seems to extend more to the northern part than to the southern part. These features can be seen more clearly in the CCH emission map. The molecular outflow is aligned along the northwest-southeast direction, and hence, the CCH emission seems to be enhanced in the outflow cavity wall, as seen in IRAS 15398-3359 and NGC1333 IRAS4C \citep{oya14, zha18, oya18b}. Therefore, the northwestern feature of c-C$_3$H$_2$ may also be affected by the outflow. 

\begin{figure}
	\centering
        \includegraphics[width = 12 cm]{\dirfig 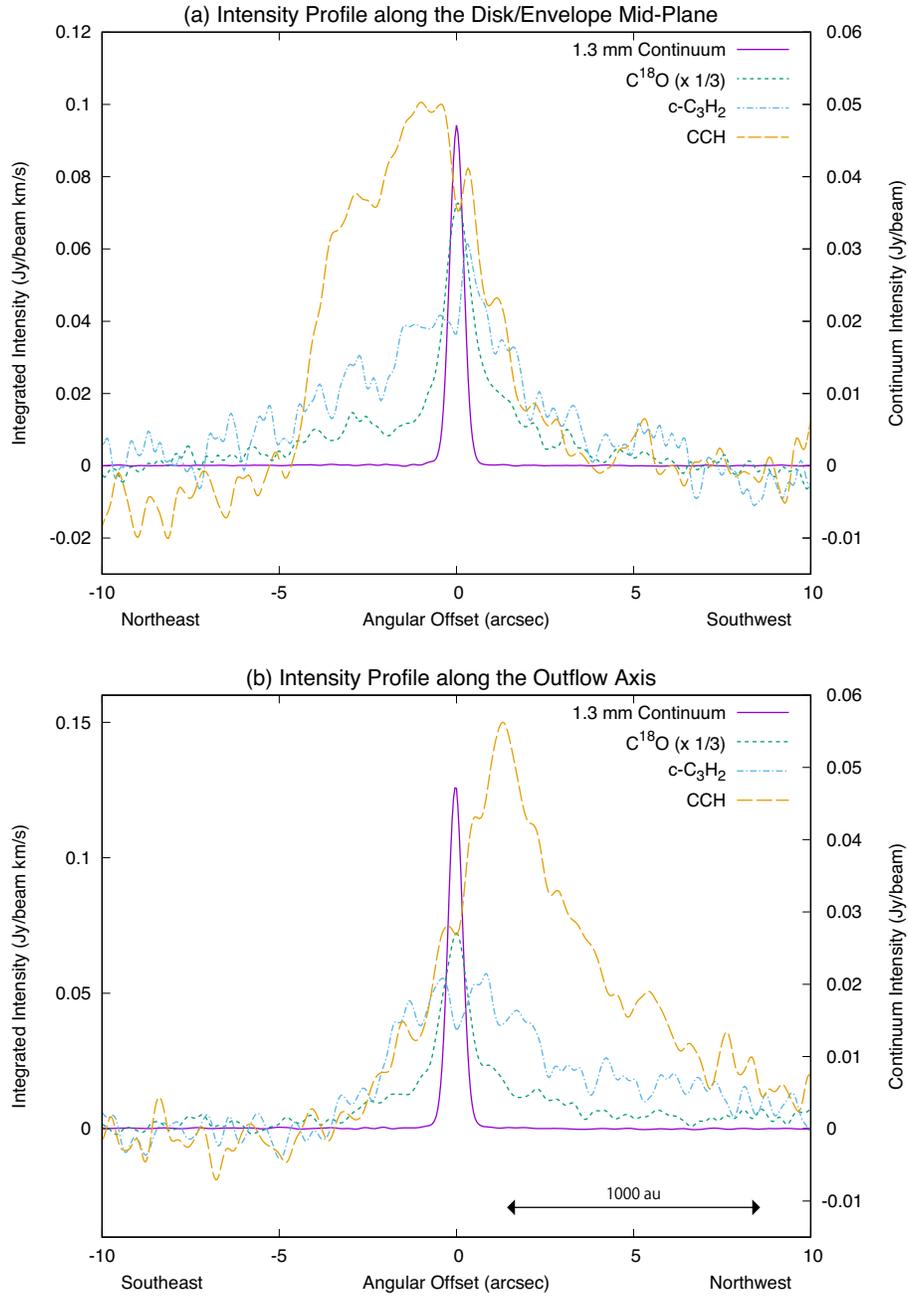} 
	\caption{(a) Intensity profiles of the C$^{18}$O ($J = 2 - 1$), c-C$_3$H$_2$ ($6_{0,6} - 5_{1,5}$; $6_{1,6} - 5_{0,5}$), CCH ($N = 3 - 2, J = 7/2 - 5/2$), and the 1.3 mm dust continuum emission 
			along the disk/envelope direction (P.A.232$^{\circ}$) indicated in the left panel of Figure~\ref{fig:CB68_moment0-ext} 
			and (b) those along the outflow axis (P.A. 142\degr). 
			The position is the offset from the continuum peak position. The rms noise level is 7 mJy beam$^{-1}$ km s$^{-1}$ for the C$^{18}$O, c-C$_3$H$_2$, and CCH lines, while it is 40 $\mu$Jy beam$^{-1}$ for the continuum emission.}
	\label{fig:CB68_intensityProfile}
\end{figure}

Figure~\ref{fig:CB68_intensityProfile}a shows the intensity profiles of the C$^{18}$O, c-C$_3$H$_2$, CCH, and continuum emission along a line passing through the continuum peak with a P.A. of 232$^{\circ}$. This direction is perpendicular to the position angle of the previously reported outflow \citep{wuu96, val00}, and hence, Figure~\ref{fig:CB68_intensityProfile}a reveals the intensity profile along the disk/envelope system. The intensities of c-C$_3$H$_2$ and CCH tend to increase approaching the protostar, although that of CCH seems heavily affected by the outflow.  Association of c-C$_3$H$_2$ and CCH with the protostar on a 1000 au scale indicates that CB68 is characterized by WCCC (e.g., \citealt{sak08}; \citealt{aik08}; \citealt{sak13}). For a reference, the intensity profiles along the outflow direction are shown in Figure~\ref{fig:CB68_intensityProfile}b. Although the distribution of c-C$_3$H$_2$ is centrally concentrated, the CCH emission shows an extension to the northwestern side along the outflow. Such extension is seen in the integrated intensity map of Figure~\ref{fig:CB68_moment0-ext}c.

\begin{figure}
	\centering
	\plotone{\dirfig 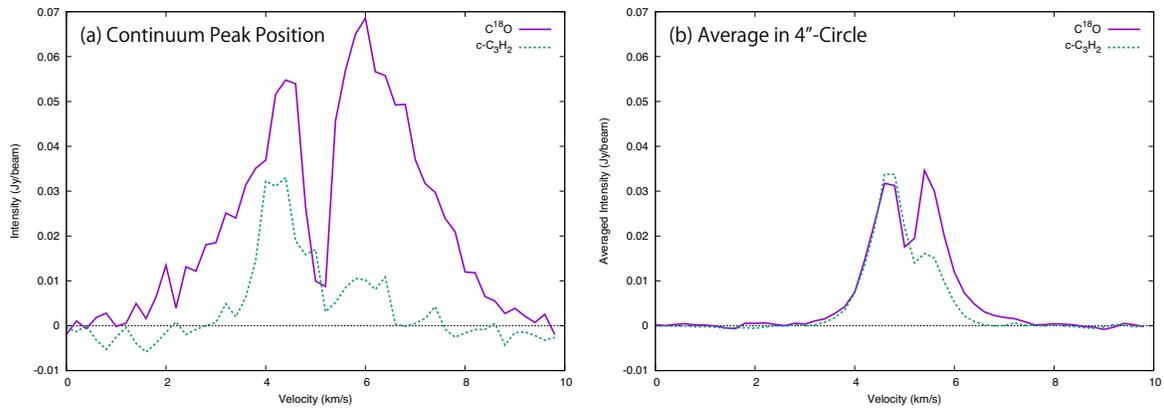}
	\caption{(a) Observed spectra of C$^{18}$O ($J = 2 - 1$) and c-C$_3$H$_2$ ($6_{0,6} - 5_{1,5}$; $6_{1,6} - 5_{0,5}$) at the continuum position. (b) Observed spectra of C$^{18}$O and c-C$_3$H$_2$ averaged over the circular area of 4\farcs0 in diameter around the continuum peak.}
	\label{fig:CB68_C18Ospectra}
\end{figure}

\textbff{The fractional abundance of c-C$_3$H$_2$ relative to H$_2$ at the continuum peak position is derived in the following way. Here, we use the C$^{18}$O emission to derive the H$_2$ column density. Figure~\ref{fig:CB68_C18Ospectra}a shows the spectra of c-C$_3$H$_2$ and C$^{18}$O at the continuum position. The spectrum of c-C$_3$H$_2$ shows self-absorption of the red-shifted component due to an infalling motion and that of the systemic velocity component due to the foreground gas (e.g., \citealt{eva15}; \citealt{yan20}). To minimize these effects, we conservatively employ the blue-shifted component, which is less affected by these effects, in the fractional abundance derivation. The peak intensities of the blue-shifted emission measured by the Gaussian fit are 32 mJy beam$^{-1}$ and 
{\bff 54 mJy beam$^{-1}$} 
for c-C$_3$H$_2$ and C$^{18}$O, respectively. From these values, the column densities of the blue-shifted component of c-C$_3$H$_2$ and C$^{18}$O are derived to be 
{\bff ($2.4-3.5) \times 10^{13}$ cm$^{-2}$ and $5.4 \times 10^{15}$ cm$^{-2}$,} 
respectively, by using the non-LTE (local thermodynamic equilibrium) radiative transfer code, RADEX. For this calculation the linewidths of 1.05 km s$^{-1}$ are used for the both species, and the H$_2$ volume densities of $10^6-10^8$ cm$^{-3}$ are assumed. This is a range of the H$_2$ density of a typical protostellar envelope \citep{sak14b}. The gas kinetic temperature is uncertain. We adopt 25 K for it, which is the desorption temperature of CH$_4$ for triggering WCCC \citep{sak08, sak13}. The column density for each species is derived to match the intensity calculated by RADEX with the observed one. The maximum optical depths of the c-C$_3$H$_2$ and C$^{18}$O lines are 
{\bff 0.55 and 0.42,} 
respectively. Then, the fractional abundance of c-C$_3$H$_2$ relative to H$_2$ is roughly estimated 
{\bff to be $(0.7-1.0) \times 10^{-9}$,} 
where the H$_2$ column density is evaluated from the C$^{18}$O column density 
{\bff to be $\sim 3.6 \times 10^{22}$ cm$^{-2}$} 
by using the empirical equation for the high C$^{18}$O column density case ($> 3 \times 10^{14}$ cm$^{-2}$) reported by \citet{fre82}. If we assume the gas kinetic temperatures of 15 K and 35 K, the fractional abundances of c-C$_3$H$_2$ 
{\bff are $(0.8-2.1) \times 10^{-9}$ and $0.6 \times 10^{-9}$,} 
respectively. In this case, the maximum optical depths of c-C$_3$H$_2$ and C$^{18}$O are 
{\bff 1.5 and 1.1,} respectively, 
for the H$_2$ density of $10^6$ cm$^{-3}$ and the temperature of 15 K. Hence, the estimated range of the fractional abundance of c-C$_3$H$_2$ 
{\bff is $(0.6-2.1) \times 10^{-9}$ in CB68,} 
which is lower by one or two orders of magnitude than the value found for L1527 ($2.7 \times 10^{-8}$) \citep{sak10}.}

\textbff{The low abundance may originate from the above analysis focused on the protostellar position with a narrow beam ($\sim 70$ au). Then, we also calculate the fractional abundance range of c-C$_3$H$_2$ averaged over the circular area of 4\farcs0 (560 au) in diameter centered at the protostar position, which is comparable to the distribution of c-C$_3$H$_2$. Figure~\ref{fig:CB68_C18Ospectra}b shows the spectra of c-C$_3$H$_2$ and C$^{18}$O averaged over the area. The self-absorption feature of the red-shifted component and the systemic velocity component is also seen as in the above case, and hence, we employ the blue-shifted component for the fractional abundance evaluation. Through the procedure mentioned above, we obtain the range of the fractional abundance of c-C$_3$H$_2$ of $(0.9-4.2) \times 10^{-9}$, which is not much different from that derived with the narrow beam. In this case, the maximum optical depths of c-C$_3$H$_2$ and C$^{18}$O are 
{\bff 1.6} and 0.5, 
respectively, for the H$_2$ density of $10^6$ cm$^{-3}$ and the temperature of 15 K. These analyses imply that the enhancement of the c-C$_3$H$_2$ abundance due to the WCCC effect is weaker in CB68 than in L1527.} 

\begin{figure}
	\centering
	\plotone{\dirfig 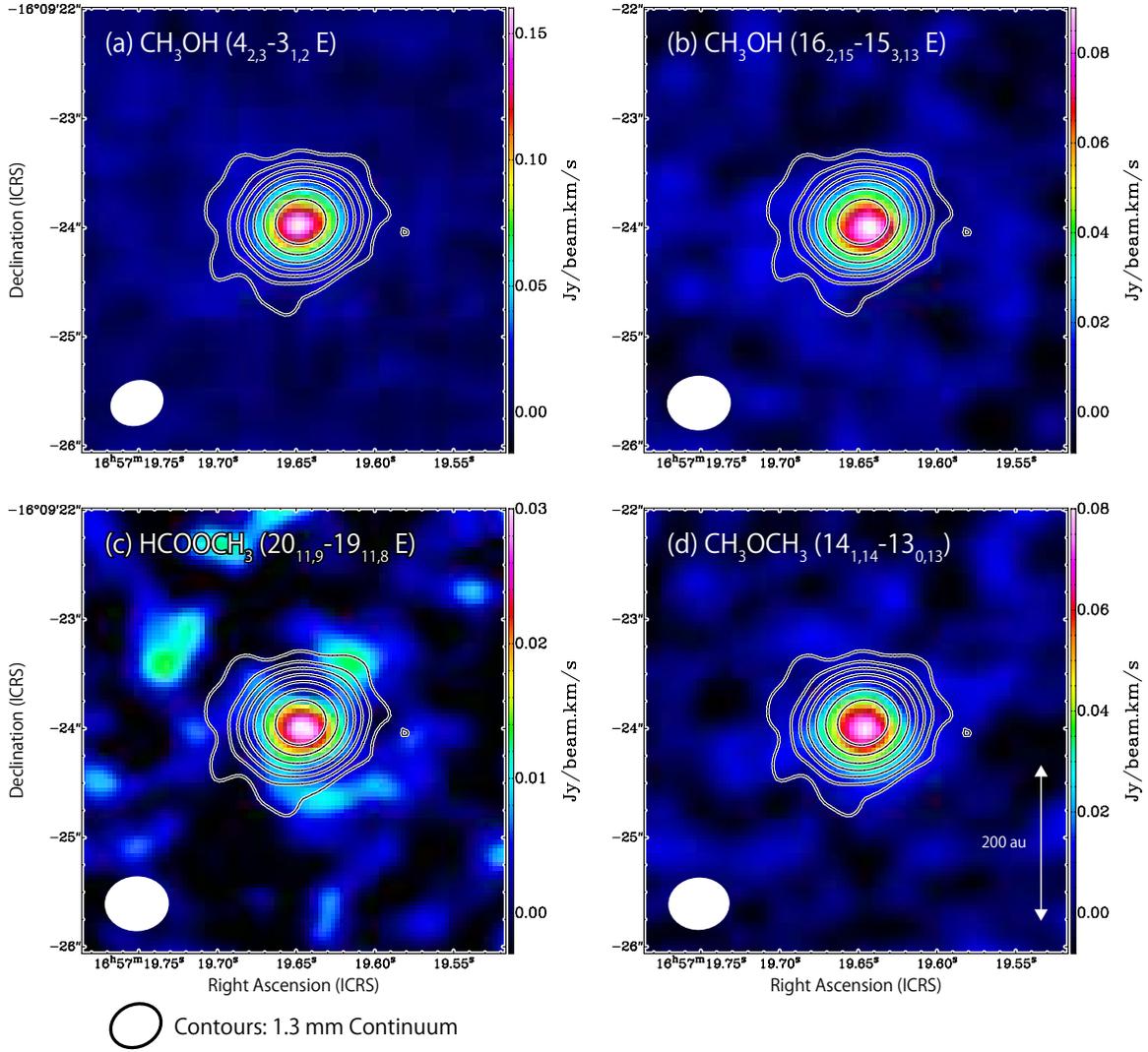} 
	\caption{Integrated intensity maps (moment 0 maps) of 
			CH$_3$OH ($4_{2,3}-3_{1,2}$, E), 
			CH$_3$OH (high excitation line; $16_{2, 15}-15_{3,13}$, E), 
			HCOOCH$_3$ ($20_{11,9}-19_{11,8}$, E), 
			and CH$_3$OCH$_3$ ($14_{1,14}-13_{0,13}$, EA, AE, EE, and AA). 
			Contours represent the continuum image of Figure~\ref{fig:CB68_cont}.  The rms noise level is 4 mJy beam$^{-1}$ km s$^{-1}$ for the CH$_3$OH ($4_{2,3}-3_{1,2}$, E) line and the HCOOCH$_3$ line and 5 mJy beam$^{-1}$ km s$^{-1}$  for the other two lines. The velocity range for the integration is from 0 to 9.8 km s$^{-1}$ for the CH$_3$OH ($4_{2,3}-3_{1,2}$, E) line and the CH$_3$OCH$_3$ line and from 0 to 10.5 km s$^{-1}$ for the other two lines. The beam sizes are shown on a bottom left corner of each panel.}
	\label{fig:CB68_moment0-COMs}
\end{figure}

On the other hand, CH$_3$OH is concentrated around the protostar on a scale of 100 au or smaller, as shown in Figures~\ref{fig:CB68_moment0-COMs}a and \ref{fig:CB68_moment0-COMs}b, 
where the low-excitation line ($4_{2,3}-3_{1,2}$ E; $E_\mathrm{u}/ k_\mathrm{B} = 45.46$ K) and the high-excitation line ($16_{2,15}-15_{3,13}$ E; $E_\mathrm{u}/k_\mathrm{B} = 338.14$ K) are depicted, respectively. 
Here, $k_\mathrm{B}$ stands for the Boltzmann constant. 
The CH$_3$OH lines show a compact distribution and are not resolved by the synthesized beam. 
In total, 11 lines of CH$_3$OH are detected (Table~\ref{tab_CB68_COM_list}).
In addition, CH$_2$DOH, HCOOCH$_3$ and CH$_3$OCH$_3$ are detected, as listed in Table~\ref{tab_CB68_COM_list}. Here, the detection criterion is that the integrated intensity is higher than the $3\sigma$ noise level.  
To identify these molecules, we visually inspect all the lines of a certain species in the observed spectral windows, confirming that line intensities are self-consistent. 
For HCOOCH$_3$, there are several faint lines in the continuum bands (Figure~\ref{fig:CB68_spectra}), 
which are reasonably explained by the calculated spectrum using the column density and the rotation temperature derived below. 
Similarly, all the intensities of the identified lines are consistent, from a posteriori check, with the derived column densities and rotation temperatures. 
Examples of the moment 0 maps of HCOOCH$_3$ and CH$_3$OCH$_3$ are shown in Figures~\ref{fig:CB68_moment0-COMs}c and \ref{fig:CB68_moment0-COMs}d, respectively. 
Their distributions are compact around the protostar and are not resolved by the synthesized beam ($\sim$ 0\farcs4$-$0\farcs5; $\sim$70$-$80~au), 
as in the case of the high-excitation line of CH$_3$OH.

We carefully check the other iCOM lines by using their simulated spectra, as applied to the HCOOCH$_3$ lines in the above paragraph. 
We notice that a line at 234471 MHz could match with the $3_{3,1}-3_{2,2}$ E line of CH$_3$CHO (234469 MHz). 
However, this line is overlapped with the CH$_2$DOH line ($8_{2,6}-8_{1,7}$ e$_0$; 234471 MHz), and moreover, the other CH$_3$CHO lines nearby this line, which are expected to be of equivalent intensity or brighter, are not detected. 
We thus conclude that CH$_3$CHO is not detected in these observations. 
Following a similar procedure, we also conclude that NH$_2$CHO, C$_2$H$_5$OH, HCOOH, CH$_3$COCH$_3$, and C$_2$H$_5$CN are non-detection (Table~\ref{tab_CB68_COM_list}). 

In CB68, the carbon-chain related species (unsaturated hydrocarbon), c-C$_3$H$_2$, is concentrated around the protostar on a scale of 1000 au, a signature of WCCC. 
In contrast, compact distributions of iCOMs are also found on a scale of 100 au or smaller, which is evidence for hot corino chemistry. 
Hence, CB68 can be regarded to have a hybrid chemistry, where WCCC and hot corino chemistries coexist but on different scales. 
Such chemical characteristics have been found in other low-mass protostellar sources such as L483 and B335 \citep{ima16, oya17}. 
It is important to emphasize that the above chemical structure found in CB68 is similar to that found in B335, which is also an isolated protostellar source. 
For quantitative comparison with the other sources in Section~\ref{sec:chemstr_compare}, we derive the molecular column densities in the following sub-section.

\subsection{Abundances of iCOMs} \label{sec:chemstr_iCOMs}
First, the column density ($N$) and the rotation temperature ($T_{\rm rot}$) of CH$_3$OH at the continuum peak are derived under the assumption of LTE. 
The emitting region for the high-excitation lines of CH$_3$OH is very compact around the protostar, 
where the density is high enough for the LTE approximation to hold.
We use the following equation that considers the effect of optical depth on the observed intensity \citep{yam17}: 
\begin{equation}
T_\mathrm{obs} = f \ \left (\frac{c^2}{2k_\mathrm{B} \nu^2} \right ) (B_\nu(T_\mathrm{rot}) - B_\nu(T_\mathrm{b}))(1 - \exp(-\tau)), \label{eq:Tobs}
\end{equation}
\noindent
and
\begin{equation}
\tau = \frac{8 \pi^3 S \mu^2}{3 h \Delta v U(T)} \left [\exp \left (\frac{h \nu}{k_\mathrm{B} T} \right ) - 1 \right ] \exp \left (\frac{-E_\mathrm{u}}{k_\mathrm{B} T} \right ) N, \label{eq:tau}
\end{equation}
where $B_{\nu}(T)$ denotes the Planck function, $U(T)$ the partition function of the molecules at the excitation temperature $T$, 
$N$ the column density, 
$\nu$ the frequency of the line, $E_\mathrm{u}$ the upper state energy, $f$ the beam filling factor, and $T_\mathrm{b}$ the cosmic microwave background temperature (2.7 K). 

\begin{sidewaysfigure}
	\centering
	\includesidewaysimg{\dirfig 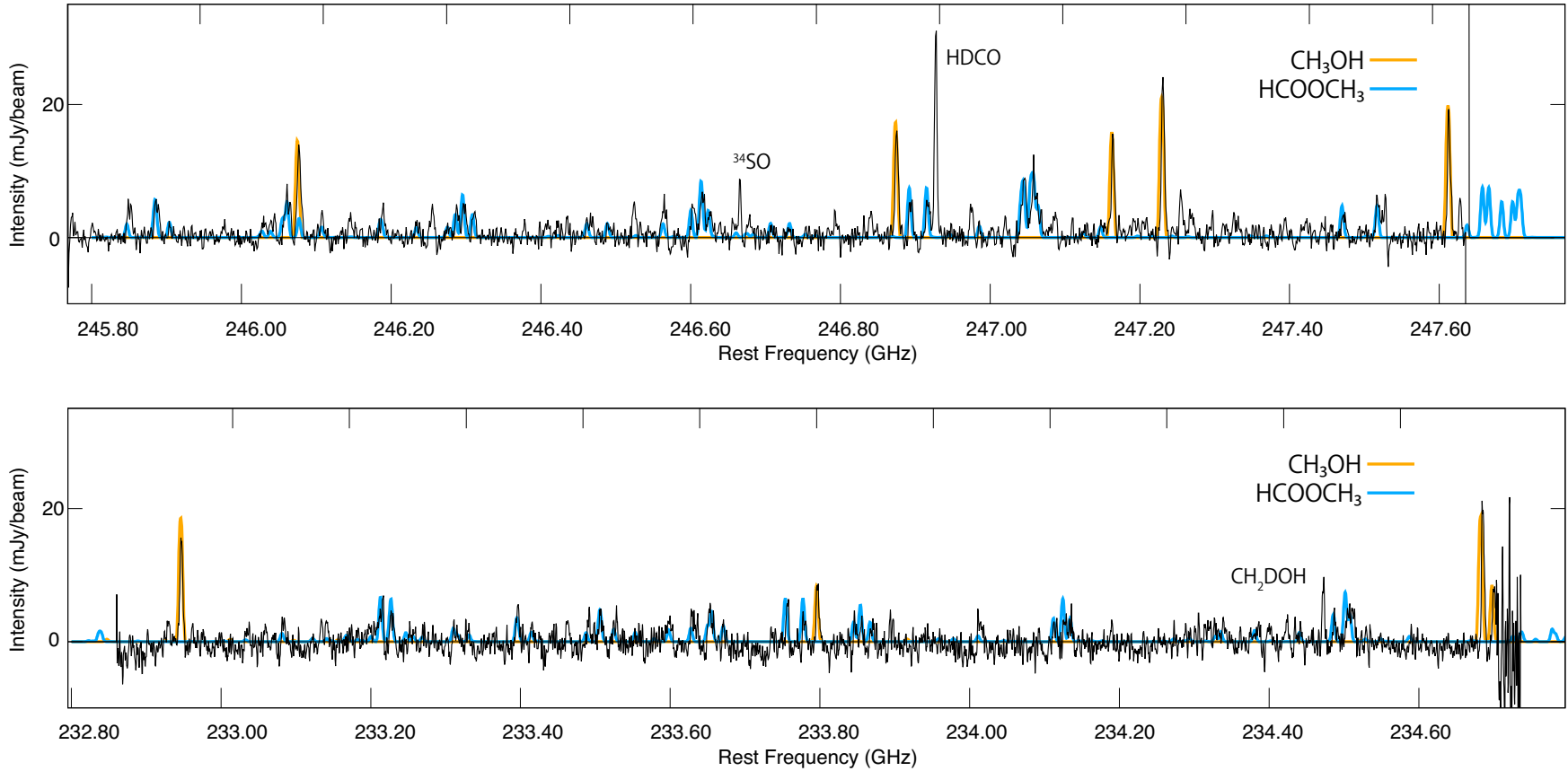} 
	\caption{Observed spectra of the two continuum bands for demonstration of rich iCOM lines (black lines). The systemic velocity of 5.0 km s$^{-1}$ is adopted to calculate the rest frequency. Colored ones are the results of the simulation based on the column density, the rotation temperature, and the beam filling factor derived in the analysis. See Section~\ref{sec:chemstr_iCOMs} for details. Orange lines represent CH$_3$OH, whereas blue ones HCOOCH$_3$.}
	\label{fig:CB68_spectra}
\end{sidewaysfigure}

The CH$_3$OH lines used in the least-squares analysis are listed in Table~\ref{tab_CB68_COM_list}. 
Among them, we find that the optical depths for the two low-excitation lines 
($4_{2,3}-3_{1,2}$ E, 218440 MHz, $E_\mathrm{u}/k_\mathrm{B}$ = 45.46 K; $5_{1,4}-4_{1,3}$ A, 243916 MHz, $E_\mathrm{u}/k_\mathrm{B}$ = 49.66 K) are higher than 6. 
Conservatively, to minimize opacity effects, we use the lines for which the opacity is less than 2. 
The rotation temperature, the column density, and the beam-filling factor are determined by the least-squares analysis \textbff{on the peak intensities which are derived by dividing the integrated intensity by the line width (6 km s$^{-1}$).} 
The largest correlation coefficient between the parameters in the least-squares fit is 0.68 between the column density and the beam filling factor. 
Hence, the parameters are well determined in the fit. 
The rotation temperature and the column density are derived to be $131\pm11$~K and $(2.7\pm1.0)\times10^{18}$~cm$^{-2}$, respectively, 
where the errors represent the standard deviation of the fit. 
The beam filling factor is well constrained and determined to be $0.022\pm0.003$. 
Since the high-excitation lines used in the analysis 
are not spatially resolved with the current synthesized beam, the small beam-filling factor is reasonable. 
This means that the CH$_3$OH emission, particularly for the high-excitation lines, mainly comes from a very compact region whose size is approximately 10 au. 
The maximum optical depth among the fitted lines is 1.9 for the $4_{2,2}-5_{1,5}$ A line. 
In Figure~\ref{fig:CB68_spectra}, 
the calculated spectrum of CH$_3$OH is overlaid on the two spectral windows. 
The observations are well-reproduced by the spectrum calculated by using the best-fit parameters. 

Similarly, the column density of HCOOCH$_3$ is derived using the least-squares fit on the observed lines listed in Table~\ref{tab_CB68_COM_list}, 
where Equation~(\ref{eq:Tobs}) is employed. Here, heavily blended lines are excluded in the analysis. 
Although we have tried to determine the column density, the rotation temperature, and the beam-filling factor simultaneously, the rotation temperature and the beam-filling factor are not well determined in the fit. This is due to a relatively poor signal-to-noise ratio of the spectra. 
Hence, we determine the column density of HCOOCH$_3$ by assuming the rotation temperature and the beam-filling factor measured for CH$_3$OH. 
This assumption is justified, because the emitting region of the high-excitation lines of CH$_3$OH and that of the HCOOCH$_3$ lines are likely to be similar (i.e., a hot corino). 
Nevertheless, we also evaluate the column density by assuming the rotation temperature of 100 K and 150 K, in order to investigate the dependence of the column density on the assumed temperature. 
The results are shown in Table~\ref{tab_CB68_COM_coldens}. 
The HCOOCH$_3$ spectrum calculated for the 131 K case is overlaid on Figure~\ref{fig:CB68_spectra}, which is consistent with the observation. 

The column densities of CH$_3$OCH$_3$ and CH$_2$DOH are derived in a similar way as HCOOCH$_3$ discussed above (Table \ref{tab_CB68_COM_coldens}). 
Specifically, the rotation temperature and the beam-filling factor are fixed to the values derived from CH$_3$OH because of small numbers of detected lines. 
The column densities are also evaluated assuming the rotation temperature of 100 K and 150 K. 
Note that the CH$_2$DOH ($4_{1,4}-4_{1,3}$ $e_1-e_0$) line at 246973 MHz was not included in the fit, because the line strength of this inter-vibrational transition is uncertain \citep{pic98}. 
Recently, the uncertainty of the line strength for this molecule has also been pointed out by \citet{amb21}. 

We derive the $3 \sigma$ upper limits to the column densities of NH$_2$CHO, CH$_3$CHO, C$_2$H$_5$OH, HCOOH, CH$_3$COCH$_3$, and C$_2$H$_5$CN, 
assuming the rotation temperature and the beam filling factor derived in the CH$_3$OH analysis. 
They are also evaluated assuming a rotation temperature of 100 K and 150 K. 
For this purpose, we simulate the spectrum of each molecule in all the spectral windows of our observations and select the highest intensity line for derivation of the upper limit. 
The upper limit to the column density is then evaluated from the $3 \sigma$ upper limit to the integrated intensity. 
The line width is assumed to be 6 km s$^{-1}$. This is the average line width of the detected iCOMs in this observation. The derived upper limits are summarized in Table~\ref{tab_CB68_COM_coldens}.

\subsection{Comparison with Other Sources} \label{sec:chemstr_compare}
In this section, we compare the molecular abundances of iCOMs in CB68 to those in several hot corinos observed with interferometers. 
To this end, column density ratios in between iCOMs are used instead of the fractional abundance relative to H$_2$, 
because the H$_2$ column density for the emitting region of iCOMs is not available in this source. 
The HCOOCH$_3$/CH$_3$OH ratio in CB68 is derived to be $0.09^{+0.07}_{-0.03}$.  
This ratio is slightly higher than values found in the other sources harboring a hot corino: 
\textbff{IRAS 16293$-$2422 Source A \citep[$0.021^{+0.011}_{-0.008}$;][]{man20},
IRAS 16293$-$2422 Source B \citep[$0.026^{+0.008}_{-0.007}$;][]{jor18}, 
NGC1333 IRAS2A \citep[$0.016^{+0.012}_{-0.007}$;][]{taq15}, 
B1-c \citep[$0.010^{+0.005}_{-0.002}$;][]{van20},
Ser-emb 8 \citep[$0.011^{+0.009}_{-0.004}$;][]{van20},
and L1551 IRS5 \citep[$0.033 \pm 0.002$;][]{bia20}.} 
The CH$_2$DOH/CH$_3$OH ratio is derived to be $0.06^{+0.04}_{-0.02}$. 
This is close to the corresponding ratios in \textbff{other hot corinos:
IRAS 16293$-$2422 Source A \citep[$0.09^{+0.04}_{-0.03}$;][]{man20},
IRAS 16293$-$2422 Source B \citep[$0.07 \pm 0.02$;][]{jor18},
B1-c \citep[$0.08^{+0.04}_{-0.02}$;][]{van20}, and
Ser-emb 8 \citep[$0.043^{+0.033}_{-0.014}$;][]{van20}.} 
It should be noted that the intrinsic line strength of CH$_2$DOH could be uncertain \citep{amb21}, so that the derived ratio may be changed by a factor of a few. 

\begin{sidewaysfigure}
	\centering
	\includesidewaysimg{\dirfig 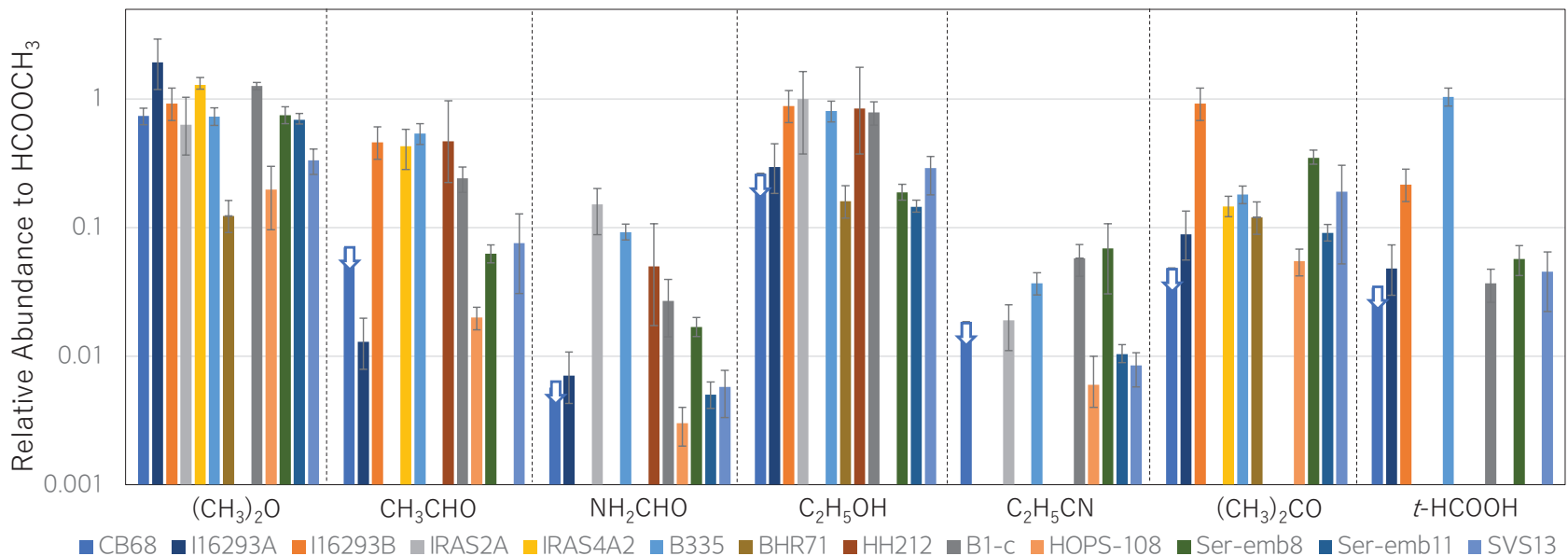}
	\caption{Comparison of relative abundances to HCOOCH$_3$ with several low-mass protostellar souces. Source names are given on the right hand side. I16393A denotes IRAS16293$-$2422 Source A, I16393B denotes IRAS16293$-$2422 Source B, IRAS2A denotes NGC1333 IRAS2A, IRAS4A2 denotes NGC1333 IRAS4A2, and BHR71 denotes BHR71 IRS1. See Table~\ref{tab_ratio-hcooch3} for references.}
	\label{fig:CB68_Comparison}	
\end{sidewaysfigure}

\begin{figure}
	\centering
	\plotone{\dirfig 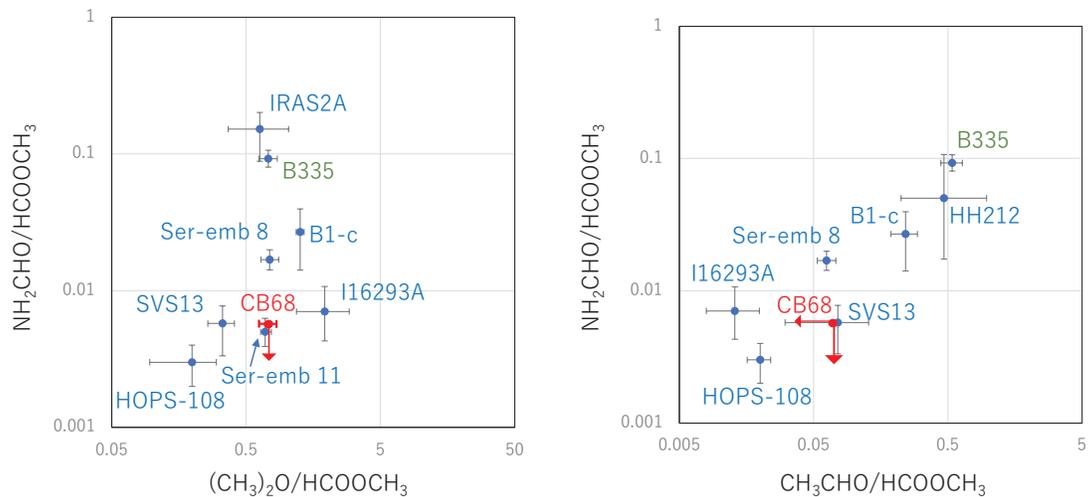} 
	\caption{(left) Plot of the abundances of CH$_3$OCH$_3$ and NH$_2$CHO relative to HCOOCH$_3$. (right) Plot of the abundances of CH$_3$OCH$_3$ and NH$_2$CHO relative to HCOOCH$_3$.}
	\label{fig:CB68_correlation}
\end{figure}

Since the CH$_3$OH abundance when considering the effect of optical depth is not always available for several hot corinos \textbff{and would have a large uncertainty}, the ratio relative to HCOOCH$_3$, which is likely optically thin in all the sources, is used for comparison to other iCOMs, as summarized in Table \ref{tab_ratio-hcooch3} and Figure~\ref{fig:CB68_Comparison}. 
\textbff{The CH$_3$OCH$_3$/HCOOCH$_3$ and C$_2$H$_5$OH/HCOOCH$_3$ ratios do not vary among the sources very much. 
Indeed}, the CH$_3$OCH$_3$/HCOOCH$_3$ ratio in CB68 is $0.74\pm0.11$, which is almost comparable to those for the other sources. 
\textbff{A trend of the constant CH$_3$OCH$_3$/HCOOCH$_3$ ratio is also reported by \citep{cha22}.}
On the other hand, \textbff{the other ratios shows some variation from source to source. Although only $3 \sigma$ upper limits to the ratio relative to HCOOCH$_3$ are obtained for the other species, the CH$_3$CHO/HCOOCH$_3$, NH$_2$CHO/HCOOCH$_3$, and HCOOH/HCOOCH$_3$ ratios in CB68 tend to be low among  the other hot corinos. 
Hence, CH$_3$CHO, NH$_2$CHO, and HCOOH seem relatively less abundant in CB68. Such a trend for CH$_3$CHO and NH$_2$CHO can also be seen in Figure~\ref{fig:CB68_correlation}.}
In particular, the difference is significant between the two isolated sources, CB68 and B335. 
Meanwhile, the upper limits to the C$_2$H$_5$OH/HCOOCH$_3$, C$_2$H$_5$CN/HCOOCH$_3$, and CH$_3$COCH$_3$/HCOOCH$_3$ ratios are comparable to the corresponding ratios in the other sources. 
Table~\ref{tab_ratio-hcooch3} lists the bolometric luminosity and the bolometric temperature of each source. 
The deficiency of CH$_3$CHO, NH$_2$CHO, and HCOOH in CB68 is not simply ascribed to these physical parameters. 

It should be noted that the dust continuum emission can be optically thick at millimeter wavelengths. 
Although the beam-averaged brightness temperature of the dust emission in CB68 is $5.73 \pm 0.07$ K, 
the dust opacity could be higher for the emitting region of CH$_3$OH. 
A detailed analysis of the dust emission is left for future study that includes analysis of the other ALMA frequency bands. 
Here, we just discuss the effect of optically thick dust on the observed molecular abundances. 
If the emitting region is the same for all the iCOMs observed in this study, 
the optically thick dust would not seriously affect the above discussions on the abundance ratios. 
The effect is compensated at least partially by taking the column density ratios. 
However, the distribution of the emission could be different from iCOM to iCOM. 
For instance, N-bearing iCOMs such as NH$_2$CHO and C$_2$H$_5$CN would be more centrally concentrated around the protostar 
than O-bearing iCOMs such as HCOOCH$_3$ and CH$_3$OCH$_3$, 
as revealed in the high-mass protostellar sources (e.g., \citealt{jim12}; \citealt{fen15}; \citealt{cse19}).
Such a trend has recently been found in the low-mass protostellar source L483 \citep{oya17, oko21}. 
If this were also the case of CB68, N-bearing iCOMs would be more affected by the optically thick dust. 
In this case, the apparent abundances of N-bearing iCOMs relative to HCOOCH$_3$ would be lower. 
Even if the dust opacity is not high, the smaller distribution yields a lower beam-averaged column density for the same beam filling factor. 
According to \citet{cse19},  
HCOOH shows a distribution like N-bearing iCOMs. 
Hence, the low abundance of HCOOH relative to HCOOCH$_3$ would also be attributed to its very compact distribution. 
In the following discussions, we need to keep these caveats in mind. 

\subsection{Comparison with Chemical Models} \label{sec:chemstr_model}
Here, we compare the observed molecular abundances with the reported results of chemical network calculations. 
\textbff{We first employ the detailed chemical model of hot cores/hot corinos with an emphasis on iCOMs reported by \citet{garr22} for comparison 
(Table~\ref{tab_iCOMs_Garrod}). This model is an updated version of that reported by \citet{gar13}. 
In those works, a three-phase model is used, 
where chemical processes in the gas-phase, on grain-surface, and in bulk-ice are fully incorporated. 
Temporal variations of molecular abundances are calculated for the three different warm-up processes, 
which mimic protostellar birth: 
namely, fast, medium, and slow warm-up processes from the cold cloud stage (10 K) to the hot corino stage (100 K). 
In the model by \citet{garr22}, the HCOOCH$_3$/CH$_3$OH ratio is 0.02 for the fast warm-up case and becomes slightly higher for slower warm-up cases. 
The model result is roughly consistent with the range of the ratio observed for the hot corino sources including CB68. 
The CH$_3$OCH$_3$/HCOOCH$_3$ ratio calculated by the model is 0.6-0.9 and does not vary significantly among the three warm-up cases. 
In the hot corino sources, 
the ratio is 0.12$-$1.9, and hence, the model result fits the observations. 
The CH$_3$CHO/HCOOCH$_3$ ratio is 0.21 for the fast warm-up case and slightly increases for slower warm-up cases. 
It is much larger than the upper limit derived in CB68, while it is consistent with some hot corinos as shown in Table \ref{tab_ratio-hcooch3} and Figure~\ref{fig:CB68_Comparison}. 
The abundance of NH$_2$CHO calculated in the fast warm-up model is 1/10 of that of HCOOCH$_3$, 
which is inconsistent with the upper limit measured for CB68 ($<0.006$). 
Note that the calculated NH$_2$CHO/HCOOCH$_3$ ratio becomes slightly higher for a slower warm-up period in the model. 
It is still higher than the observed ratio for the most of sources except for NGC1333 IRAS2A.}  

On the other hand, \citet{aik20} presented chemical network calculations considering the physical evolutions of the static and collapsing phases, 
where the physical condition is taken from the 1D radiation hydrodynamic model of low-mass star formation presented in \citet{mas98} and \citet{mas00}. 
Their results successfully reproduce both of the chemical features observed in CB68: (1) the enhancement of carbon-chain molecules and related species on a 1000 au scale (i.e., WCCC) and (2) enrichment of iCOMs in the vicinity of the protostar (i.e., hot corino chemistry).
The chemical network is essentially the same as that reported by \citet{gar13} except for some updates of the reactions. 
However, the detailes of the ice chemistry model, i.e., multi-layered ice mantle without swapping, are different from \citet{gar13}, which affect the iCOM abundances.

According to 
\citet{aik20}, 
the HCOOCH$_3$/CH$_3$OH and CH$_3$OCH$_3$/HCOOCH$_3$ abundance ratios \textbff{in the hot corino stage} derived from the chemical network calculation is much lower and higher, respectively, than the observational results for the hot corino sources including CB68. 
\citet{aik20} also reported that the CH$_3$OCH$_3$/HCOOCH$_3$ ratio depends on \textbff{the lowest temperature $T_{\rm{min}}$ during the evolution. The ratio changes by two orders of magnitude, where it is close to unity for \textbff{the case of $T_{\rm{min}}$ of 20 K: the ratio is higher for the lower and higher $T_{\rm{min}}$ in the range from 10 K to 25 K.} For $T_{\rm{min}}$ of 20 K, the CH$_3$OCH$_3$/HCOOCH$_3$ ratio is almost comparable to the observed ratios in hot corinos, while the CH$_3$CHO/HCOOCH$_3$ and NH$_2$CHO ratios are much overestimated. The HCOOCH$_3$ abundance seems much underestimated, and resolving this issue is left for a future study.} 

\section{Kinematic Structure} \label{sec:kin} 
In this section, we explore the kinematic structure of the hot corino and its surrounding envelope. 
We assume the P.A. of the disk/envelope system to be 232\degr, perpendicular to the outflow direction (P.A. = 142\degr) \citep{wuu96,val00,lau10}.
The disk/envelope direction is shown by dashed lines of Figure~\ref{fig:CB68_moment0-ext}a. 

\begin{figure}
	\centering
	\plotone{\dirfig 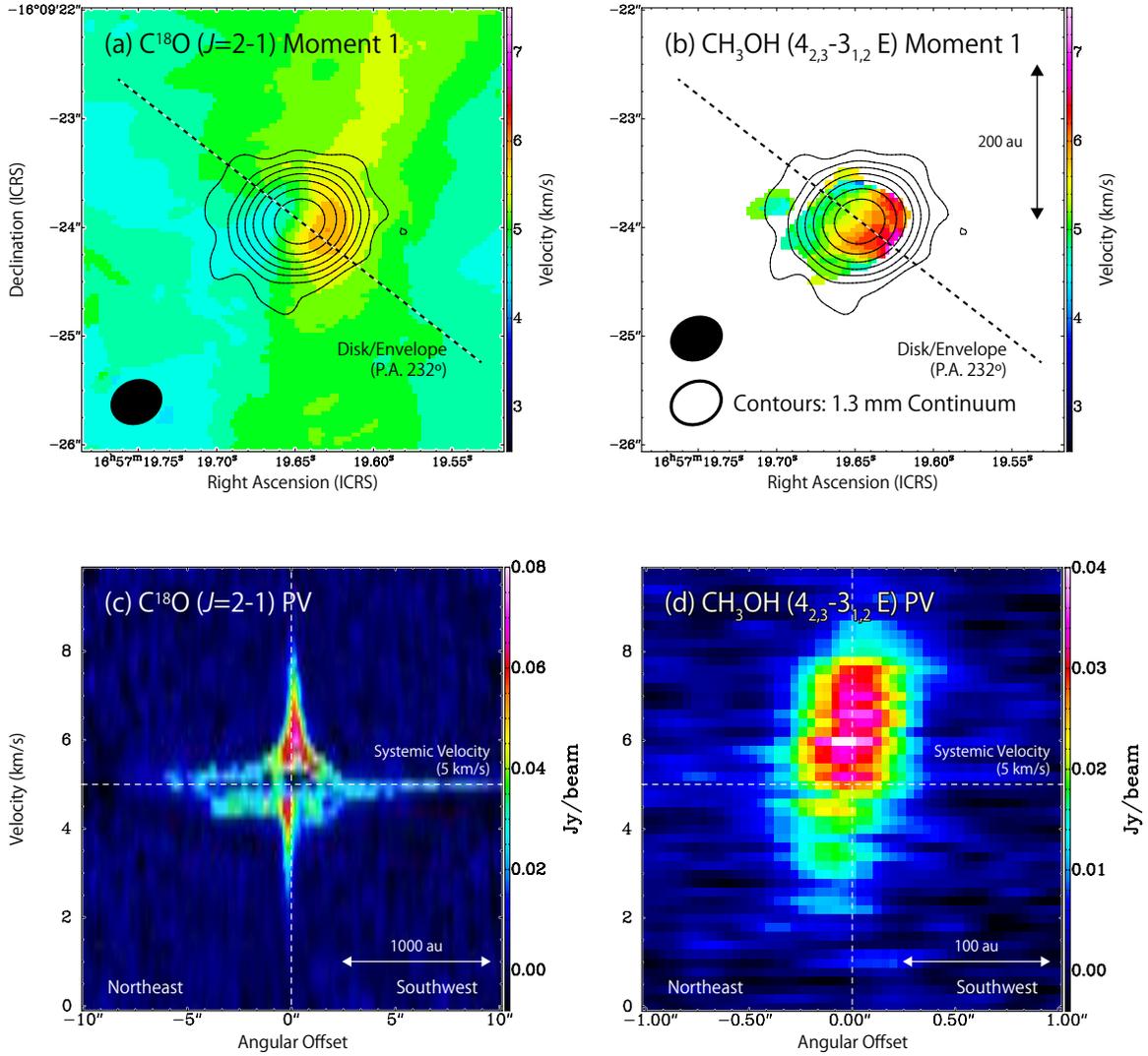} 
	\caption{(a, b) The velocity field maps (moment 1 maps) of C$^{18}$O and CH$_3$OH. Black closed ellipses represent the beam sizes. The cube data with the intensity less than 3$\sigma$ are not used. Furthermore, the positions with the integrated intensity less than 3$\sigma$ are shown in white. Contours represent the continuum image of Figure~\ref{fig:CB68_cont}, where a black open ellipse shows its beam size. Dashed lines represent the disk/envelope direction, along which the position velocity diagrams are presented.  (c) The position velocity diagram of C$^{18}$O along the disk/envelope direction shown in Figure~\ref{fig:CB68_moment0-ext}. (d) The position velocity diagram of CH$_3$OH along the disk/envelope direction. A smaller range of the position offset is employed to see the velocity gradient clearly.}
	\label{fig:CB68_mom1-PV}
\end{figure}

\begin{figure}
	\centering
	\plotone{\dirfig 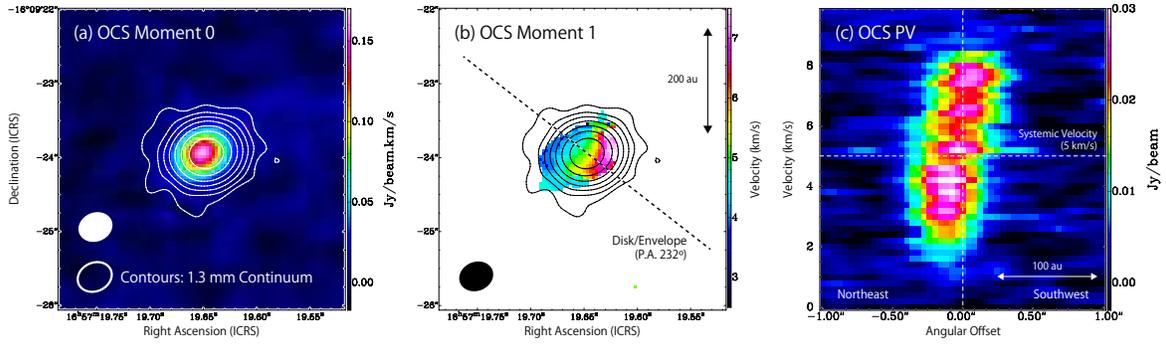} 
	\caption{The integrated intensity map (left), the moment 1 map (middle), and the PV diagram of OCS along the disk/envelope direction (right), where a dashed line in the middle panel shows the disk/envelope direction, The velocity range for the integrated intensity map is from 0.0 km s$^{-1}$ to 9.8 km s$^{-1}$. Contours on the left and middle panels represent the continuum emission (Figure~\ref{fig:CB68_cont}). See the caption of Figure~\ref{fig:CB68_mom1-PV} for preparation of the moment 1 map. The rms noise level of the integrated intensity map is 4 mJy beam$^{-1}$ km s$^{-1}$. On the left panel, a white closed ellipse represents the beam size of the OCS image, while a white open ellipse that of the 1.3 mm continuum image. On the middle panel, a black closed ellipse represents the beam size of the OCS image. A velocity gradient is visible in the moment 1 map and the PV map.}
	\label{fig:CB68_OCS}
\end{figure}

Figures~\ref{fig:CB68_mom1-PV}a and \ref{fig:CB68_mom1-PV}b show the moment 1 maps of the C$^{18}$O ($J = 2 - 1$) line and the low-excitation line ($4_{2,3} - 3_{1,2}$ E) of CH$_3$OH. 
A velocity gradient along the disk/envelope direction is seen for these two lines, although the gradient in the moment 1 map of C$^{18}$O looks marginal due to the contribution of the extended component. 
To investigate the velocity structure in more detail, we draw position-velocity (PV) diagrams along the disk/envelope direction. 
Figure~\ref{fig:CB68_mom1-PV}c shows the PV diagram of the C$^{18}$O line along the disk/envelope direction. 
A diamond shape feature can be seen, where the velocity width increases approaching the protostar. 
Moreover, a marginal velocity gradient, suggesting a rotation motion, is found in the vicinity of the protostar: 
the southwestern side is red-shifted, while the northeastern side is blue-shifted. 
This velocity gradient is more clearly seen in the PV diagram of the low-excitation CH$_3$OH line observed with the high frequency-resolution spectral window (Figure~\ref{fig:CB68_mom1-PV}d). 
Although the blue-shifted emission is rather faint, the velocity gradient is consistent with that found in C$^{18}$O. 
For further confirmation, we investigate the velocity structure of the OCS ($J = 19 - 18$) emission, which is concentrated around the protostar (Figure~\ref{fig:CB68_OCS}a).
Figure~\ref{fig:CB68_OCS}b shows the moment 1 map, while Figure~\ref{fig:CB68_OCS}c depicts the PV diagram along the disk/envelope direction.  
OCS is also reported to be enhanced in the inner envelope of IRAS 16293$-$2422 Source A as well as iCOMs \citep{oya16}. 
The velocity gradient ($\sim 0.07$ km s$^{-1}$ au$^{-1}$) is clearly seen in the moment 1 map and the PV diagram of OCS. 

\begin{figure}
	\centering
	\plotone{\dirfig 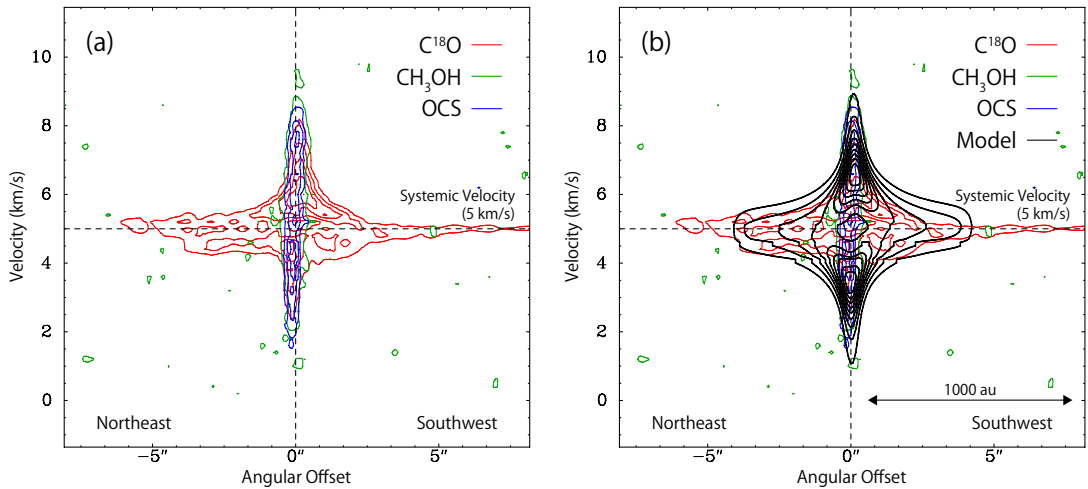} 
	\caption{(a) A composite PV diagram of C$^{18}$O (red), CH$_3$OH (green), and OCS (blue), which represents the kinematic structure for a wide spatial range. 
			(b) Same as (a) with the result of the infalling rotating envelope model using the fiducial values of the model parameters (black contours). 
			See Section \ref{sec:kin} for details. 
			The position shows the offset from the continuum peak.} 
	\label{fig:CB68_PV_model_rep} 
\end{figure}

The diamond shape structure seen in the PV diagram of C$^{18}$O is a typical feature characteristic of an infalling-rotating envelope (IRE), 
where the gas is infalling with conserved angular momentum 
\citep{oha97, sak14, oya14, oya16}. 
The PV diagram further shows a counter velocity component (blue-shifted emission in the southwestern part and red-shifted emission in the northeastern part). 
Such feature is difficult to explain by a Keplerian motion alone. 
To compare the kinematic structures of the above molecules, we present a composite PV diagram, where the individual PV diagrams of the C$^{18}$O, CH$_3$OH, and OCS lines are overlaid in the same panel (Figure~\ref{fig:CB68_PV_model_rep}a). 
Although the distribution of the CH$_3$OH line is not well resolved, 
its kinematic structure resembles that of the OCS line. 
While C$^{18}$O seems to trace an entire IRE, CH$_3$OH and OCS most likely trace its innermost part.

Although the velocity gradient is small, we try to roughly estimate physical parameters of the IRE. 
For this purpose, we compare the kinematic structures observed for the C$^{18}$O, CH$_3$OH, and OCS lines with that derived from a simple model of IREs. Details of the model are described in \citet{oya14}. 
This model has successfully been applied to the IREs of some low-mass protostellar sources (e.g., \citealt{sak14}; \citealt{oya15}; \citealt{oya16}; \citealt{oya17}; \citealt{ima19}; \citealt{oya20}). 
In this model, the structure of the IRE is determined by six free parameters:  the kinematic structure is specified by (1) the radius of the centrifugal barrier ($r_\mathrm{CB}$) and (2) the mass of the central protostar ($M_\ast$), while the shape of the IRE is specified by (3) the outer radius of the emitting region in the envelope ($R_\mathrm{0}$), (4) the full thickness of the envelope ($h_\mathrm{0}$), (5) the flare angle of the envelope, and (6) the inclination angle ($i$) with respect to the line of sight ($i=90^{\circ}$ for the edge-on case). 

In the CB68 model, we make the following assumptions. 
The outer radius is fixed to be 600 au by referring the distribution of the C$^{18}$O emission. 
We assume the inclination angle to be 70\degr \ (nearly edge-on disk). This assumption seems reasonable because the well-collimated outflow feature \citep{val00} suggests the nearly edge-on geometry of the disk. 
Moreover, the fitting results do not depend on inclination angles ranging from 50\degr \ to 90\degr, although the protostellar mass is correlated with it ($\propto sin^{-2}i$).  
We arbitrarily assume a flare angle of the envelope of 10\degr \ and that the full thickness of the envelope is 100 au. 
We confirm that the results do not significantly change for a flare angle between 0\degr \ and 30\degr \ and for a thickness between 10 au to 100 au.
This is because the resolution of the present observation ($\sim$0\farcs4$-$0\farcs5; 70$-$80~au) is insufficient to place strong constraints on these parameters.

\begin{sidewaysfigure}
	\centering
	\includesidewaysimg{\dirfig 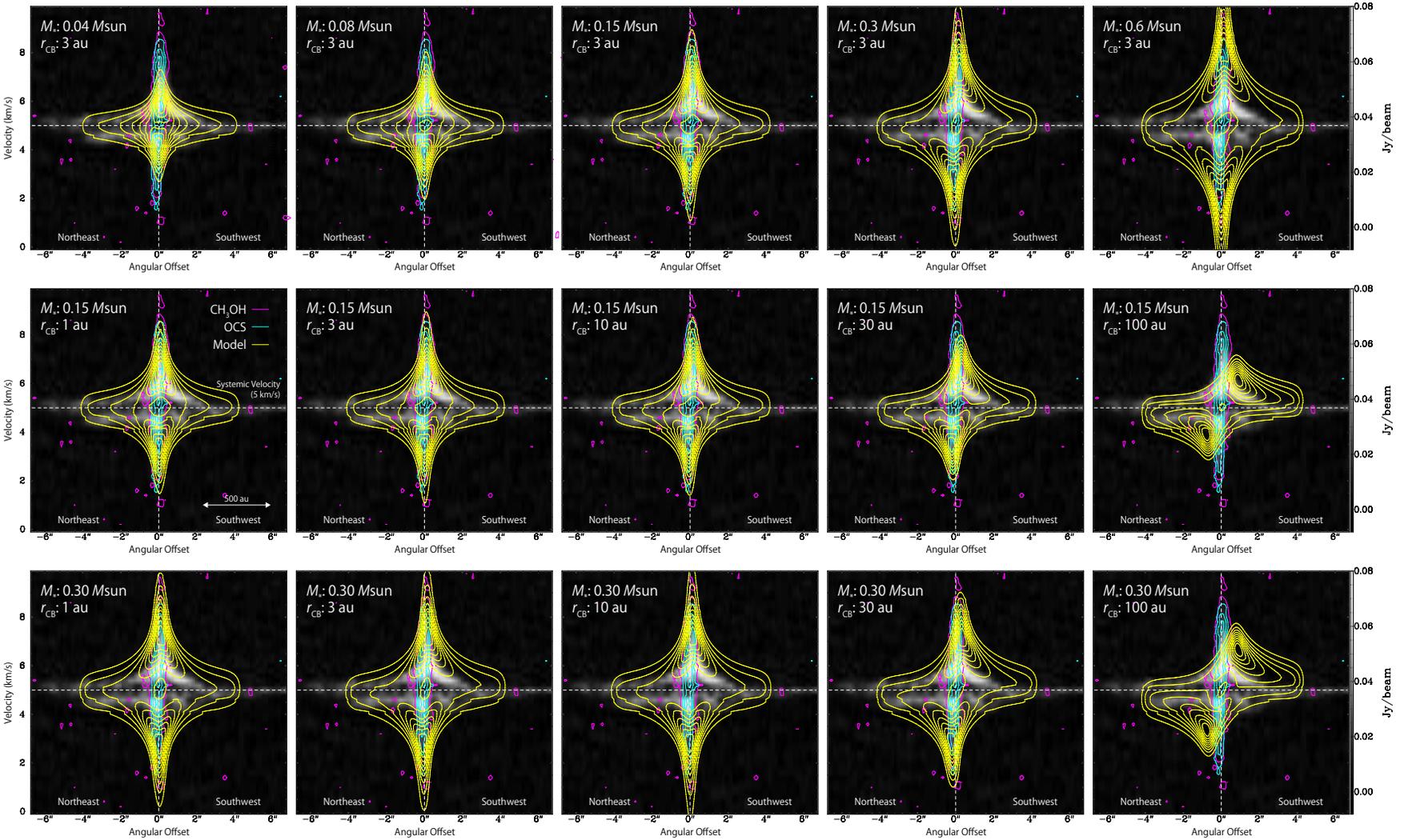} 
	\caption{Results of the infalling rotating envelope model with various parameters (yellow contours) 
			superposed on the composite PV diagram shown in Figure~\ref{fig:CB68_PV_model_rep}a. C$^{18}$O is presented in gray scales, CH$_3$OH in magenta, and OCS in cyan. The position shows the offset from the continuum peak. 
			The protostellar mass of 0.08$-$0.30 $M_\odot$ and the radius of the centrifugal barrier ($r_{\rm CB}$) of 30 au or smaller 
			can reasonably reproduce the observed kinematic structure. }
	\label{fig:CB68_PV_paravar}
\end{sidewaysfigure}

Figure~\ref{fig:CB68_PV_model_rep}b shows the result of the model calculation overlaid on the composite PV diagram. 
The velocity structure including the counter velocity component seems to be reproduced with 
 $r_\mathrm{CB}$ of 3 au and $M_\ast$ of 0.15 $M_\odot$. 
We use these values as the fiducial values to find the acceptable ranges of the parameters. 
To estimate the ranges of $r_\mathrm{CB}$ and $M_\ast$, we change their values from the above fiducial ones and compare the results with the composite PV diagram by eye. Figure~\ref{fig:CB68_PV_paravar} shows a comparison with the composite PV diagram. In the comparison, we focus on the overall velocity extent rather than the detailed intensity distribution because the latter is difficult to be reproduced quantitatively due to various factors such as the self-absorption effect and inhomogeneous distribution not considered in the IRE model.   
The velocity shift is mainly affected by $M_\ast$ because the velocity of the gas at a certain radius is roughly proportional to $M_\ast ^{1/2}$ in the IRE model. 
Considering that the OCS and CH$_3$OH emission has higher velocity component than the C$^{18}$O emission around the protostar position, we conservatively adopt the ranges from 0.08 $M_\odot$ to 0.30 $M_\odot$ for  $M_\ast$. 
On the other hand, $r_\mathrm{CB}$ affects the velocity gradient in the PV diagram. 
The observed PV diagrams can be reasonably explained by $r_\mathrm{CB}$ less than 30 au \textbff{for the models assuming $M_\ast$ of 0.15 $M_\odot$. This result is not changed if $M_\ast$ is changed in the above range. For instance, the case for 
{\bff 0.3 $M_\odot$} 
is shown in Figure~\ref{fig:CB68_PV_paravar} (bottom panel).}  

\section{Comparison with B335} \label{sec:B335}
In this section, we discuss the similarities and differences between CB68 and B335, \textbff{both of which} are isolated Class 0 protostellar sources in Bok globules and have similar bolometric luminosity and bolometric temperature (Table~\ref{tab_ratio-hcooch3}).
\textbff{Note that BHR71 IRS1 is also a Class 0 protostar in a Bok globule, whose chemical compositions may be compared with those of CB68 in the context of the isolated protostars. However, this source accompanies another protostar IRS2 separated from IRS1 by 16\arcsec, and hence, it is not in a fully isolated condition like CB68 and B335. Furthermore, the abundances are obtained for a limited species, and the data of NH$_2$CHO and CH$_3$CHO, which show relatively low abundances in CB68 are not available \citep{yan20}. For these reasons, we do not particularly discuss BHR71 IRS1 for detailed comparison with CB68 in this paper.}

The protostellar mass of CB68 estimated in this study is likely higher than the mass of B335 \citep[0.02$-$0.06~$M_\odot$;][]{ima19}. 
Only the upper limit to the radius of the centrifugal barrier is obtained for the both sources ($<$ 5 au for B335 and $<$ 30 au for CB68). 
These radii are smaller than those previously reported for the other protostellar sources L1527 \citep[$100 \pm 20$ au:][]{sak14b}, 
IRAS 16293$-$2422 Source A \citep[40$-$60 au:][]{oya16},
and TMC-1A \citep[$\sim 50$ au:][]{sak16}.
The comparatively small centrifugal barrier indicates a relatively small specific angular momentum of the accreting gas and that 
the infalling motion is dominant in the protostellar envelope. Hence, the Keplerian disk has not grown to a sufficient size to be detected inside the centrifugal barrier. 
Although we cannot resolve structure finer than 30 au in this study, a small emitting region of CH$_3$OH indicates that the size of the hot corino is about 10 au. 
If iCOMs are distributed in the transition zone from the envelope to the disk as reported for IRAS 16293$-$2422 Source A and B335 \citep{oya16, ima19}, 
the small emitting region of CH$_3$OH is consistent with the small radius of the centrifugal barrier. 
In this case, a weak accretion shock would be responsible for liberation of iCOMs around the centrifugal barrier \citep{sak14b, oya16, gar22}. 
Although the disk size of Class 0 protostellar sources is reported to be generally small \citep{mau19}, the very small radius for CB68 and B335 might be related to their isolated condition. 
For instance, it is suggested that higher exposure to the interstellar radiation field can increase the local ionization fraction which would produces small disk structures (e.g., \citealt{zhao18}; \citealt{kue20}). However, we should note that this suggestion is still controversial \citep{zhao20}.

Given the physical conditions detailed above, we find that the overall chemical structure of CB68 resembles that of B335. 
CB68 is a hybrid type consisting of the WCCC on a 1000 au scale and the hot corino chemistry on a few 10 au scale around the protostar. 
Such a feature is consistent with the result of the chemical network calculation by \citet{aik20}, as mentioned in Section~\ref{sec:chemstr_model}. 
This result seems consistent with the idea that the parent cores of the isolated sources like CB68 and B335 are well exposed to the interstellar radiation field, which may prevent the total conversion of C into CO and subsequent iCOM formation on dust grains. 
On the other hand, unsaturated hydrocarbons are produced efficiently in the gas phase. 
In a small central part where the extinction is large enough, the C to CO conversion efficiently occurs and iCOM formation on dust grains as well. 

\textbff{Although the isolated sources CB68 and B335 both show the hybrid chemical nature, we should note that the hybrid chemical nature is not specific to isolated sources. Indeed, the hybrid chemistry is reported for L483 \citep{oya17}, and its hint is also seen in Serpens SMM3 \citep{tyc21}. Nevertheless, simple physical and chemical structures of isolated sources can be used as the best testbed to study chemical evolution during the protostellar evolution without influences of nearby young stellar objects.}

Despite such similarity between CB68 and B335, there is one notable relative chemical difference between the two sources. 
Namely, CH$_3$CHO, NH$_2$CHO, and HCOOH are less abundant in CB68 than in B335 (Table \ref{tab_ratio-hcooch3}). 
This difference remains unexplained. 
Although it could be ascribed to the different local density/temperature, 
we have yet to consider the effect of the dust opacity in the iCOM emitting region before further investigating the chemistry. 
The FAUST project will further explore other protostellar sources at 50 au resolution with equivalent sensitivity and will provide the additional observations necessary to address this problem.

\section{Summary} \label{sec:summary}

The 1.3~mm band observations of CB68 were conducted with ALMA to characterize the chemical and physical structures of the protostellar envelope. The main results are summarized below.

\begin{enumerate}
\item Carbon-chain related molecules CCH and c-C$_3$H$_2$ as well as C$^{18}$O are detected in the envelope on a scale of 1000 au. 
The overall distribution of the c-C$_3$H$_2$ emission is concentrated around the continuum peak, indicating the WCCC nature of this source. 
The CCH emission seems to trace the outflow cavity extending to the northwestern direction, as well. 
\item CH$_3$OH, CH$_2$DOH, HCOOCH$_3$, and CH$_3$OCH$_3$ are detected in the vicinity of the protostar on a scale less than the beam size ($\sim$70-80~au). 
Detection of iCOMs clearly indicates that CB68 harbors a hot corino. 
The rotation temperature of CH$_3$OH is derived to be $131\pm11$~K by assuming the LTE condition. 
The emitting region of CH$_3$OH is as small as 10 au. 
\item The HCOOCH$_3$/CH$_3$OH and CH$_3$OCH$_3$/HCOOCH$_3$ ratios obtained in this source are comparable to those in the other hot corino sources and B335. 
On the other hand, CH$_3$CHO, NH$_2$CHO, and HCOOH are less abundant than in the other sources. 
\item The hybrid-type chemical characteristics (hot corino chemistry and WCCC) is thus revealed in CB68, which resembles the case of another isolated protostellar source B335.
\item  The whole envelope structure is traced by C$^{18}$O, while only the innermost part of the envelope is traced by CH$_3$OH and OCS. 
These molecules show a marginal velocity gradient perpendicular to the outflow (disk/envelope direction). 
The position velocity diagram along the disk/envelope direction is explained by the IRE model. 
The protostellar mass and the radius of the centrifugal barrier are estimated to be \textbff{$0.08-0.30$~$M_\odot$} and $< 30$ au, respectively. 
A small centrifugal barrier may be related to the small emitting region of iCOMs, if the iCOM emission mainly originates from the transition zone around the centrifugal barrier. \\
\end{enumerate}

\clearpage

\acknowledgments
The authors thank the anonymous reviewer for valuable comments and suggestions. This paper makes use of the following ALMA data: ADS/JAO.ALMA\#2016.1.01376.S. ALMA is a partnership of ESO (representing its member states), NSF (USA) and NINS (Japan), together with NRC (Canada), NSC and ASIAA (Taiwan), and KASI (Republic of Korea), in cooperation with the Republic of Chile. The Joint ALMA Observatory is  operated by ESO, AUI/NRAO and NAOJ. 
The authors acknowledge the financial support by JSPS and MAEE under the Japan-France integrated action programme (SAKURA: 25765VC). 
D.J. is supported by NRC Canada and by an NSERC Discovery Grant. 
EB and GB acknowledge funding from the European Research Council (ERC) under the European Union's Horizion 2020 research and innovation programme, for the Project ''The Dawn of Organic Chemistry'' (DOC), grant agreement No 741002.
G.B. acknowledges support from the PID2020-117710GB-I00 grant funded by MCIN/ AEI /10.13039/501100011033 and from the ``Unit of Excellence Mar\'ia de Maeztu 2020-2023 award to the Institute of Cosmos Sciences (CEX2019-00918-M).
I.J.-S. has received partial support from the Spanish State Research Agency (AEI) through project numbers PID2019-105552RB-C41 and MDM-2017-0737 Unidad de Excelencia ''Maria de Maeztu'' - Centro de Astrobiologia (CSIC-INTA).
This study is supported by Grant-in-Aids from Ministry of Education, Culture, Sports, Science, and Technologies of Japan (18H05222, 18J11010, 19H05069, 19K14753, and 21K13954).



\clearpage

\clearpage 

\clearpage 

\startlongtable
\scriptsize
\begin{deluxetable}{lccccccc}
\tablecaption{Parameters for Used Lines 
	\label{tab_CB68_COM_list}} 
\tablehead{
\colhead{Molecule} & \colhead{Transition} & \colhead{Frequency} & \colhead{$S\mu^2$} & \colhead{$E_{\rm u}$} & \colhead{$W$\tablenotemark{a}} & \colhead{maj\tablenotemark{b}} & \colhead{min\tablenotemark{b}} \\ 
& & \colhead{MHz} & \colhead{Debye$^2$} & \colhead{K} & \colhead{Jy beam$^{-1}$ km s$^{-1}$} & \colhead{arcsec} & \colhead{arcsec} 
}
\startdata
C$^{18}$O        & $J = 2-1$                      & 219560.3541 &  0.0244  &  15.8060 & 0.218(7) & 0.491 & 0.411 \\
c-C$_3$H$_2$     & $6_{0,6}-5_{1,5}$            & 217822.1480 &  58.3  &  38.6077 & 0.037(6)\tablenotemark{c} & 0.502 & 0.411 \\
                 & $6_{1,6}-5_{0,5}$            & 217822.1480 & 175  &          &          &       &       \\	
CCH              & $N$=3-2 $J$=7/2-5/2 $F$=4-3  & 262004.2600 &   2.29   &  25.1494 & 0.071(7)\tablenotemark{c} & 0.551 & 0.476 \\
                 & $N$=3-2 $J$=7/2-5/2 $F$=3-2  & 262006.4820 &   1.71   &  25.1482 &          &       &       \\ 
CH$_3$OH            & $4_{2,3}-3_{1,2}$ E           & 218440.0630 &  3.47    &  45.4600 & 0.162(6) & 0.497 & 0.409 \\
                 & $5_{1,4}-4_{1,3}$ A           & 243915.7880 &  3.88   &  49.6607 & 0.209(6) & 0.591 & 0.506 \\ 
                 & $10_{3,7}-11_{2,9}$ E         & 232945.7970 &   3.04   & 190.3713 & 0.091(6) & 0.470 &.0.388 \\
                 & $18_{3,15}-17_{4,14}$ A       & 233795.6660 &  5.47   & 446.5851 & 0.051(7) & 0.469 & 0.388 \\
                 & $4_{2,3}-5_{1,4}$ A           & 234683.3700 &   1.12   &   60.9237 & 0.117(6) & 0.467 & 0.387 \\
                 & $5_{4,2}-6_{3,3}$ E           & 234698.5190 &   0.46   &  122.7230 & 0.047(6) & 0.467 & 0.387 \\
                 & $20_{3,17}-20_{2,18}$ A       & 246074.6050 &  19.5   & 537.0395 & 0.098(5) & 0.587 & 0.503 \\
                 & $19_{3,16}-19_{2,17}$ A       & 246873.3010 &  18.4   & 490.6542 & 0.102(6) & 0.586 & 0.502 \\
                 & $16_{2,15}-15_{3,13}$ E       & 247161.9500 &  4.83   & 338.1423 & 0.091(7) & 0.585 & 0.501 \\ 
                 & $4_{2,2}-5_{1,5}$ A           & 247228.5870 &   1.09   &  60.9253 & 0.144(6) & 0.584 & 0.500 \\
                 & $18_{3,15}-18_{2,16}$ A       & 247610.9180 &  17.4   & 446.5852 & 0.116(7) & 0.584 & 0.500 \\
HCOOCH$_3$       & $19_{4,16}-18_{4,15}$ A       & 233226.7880 &  48.0   &  123.2487 & 0.030(6) & 0.470 & 0.388 \\ 
                 & $18_{4,14}-17_{4,13}$ E       & 233753.9600 &  45.8   &  114.3711 & 0.028(7) & 0.468 & 0.387 \\ 
                 & $18_{4,14}-17_{4,13}$ A       & 233777.5210 &  45.8   &  114.3651 & 0.021(7) & 0.469 & 0.388 \\
                 & $20_{13,7}-19_{13,6}$ E       & 245883.1790 &  92.5   & 235.9818 & 0.047(6)\tablenotemark{c} & 0.588 & 0.504 \\
                 & $20_{13,7}-19_{13,6}$ A       & 245885.2430 &          &          &          &       &       \\
                 & $20_{13,8}-19_{13,7}$ A       & 245885.2430 &          &          &          &       &       \\
                 & $20_{11,9}-19_{11,8}$ E       & 246285.4000 &  37.2   & 204.2141 & 0.031(6) & 0.587 & 0.503 \\ 
                 & $20_{11,10}-19_{11,9}$ A      & 246295.1350 &  74.4   & 204.2103 & 0.039(6)\tablenotemark{c} & 0.586 & 0.502 \\ 
                 & $20_{11,9}-19_{11,8}$ A       & 246295.1350 &          &          &          &       &       \\
                 & $20_{11,10}-19_{11,9}$ E      & 246308.2720 &  37.2   & 204.2014 & 0.027(6) & 0.586 & 0.502 \\ 
                 & $20_{10,10}-19_{10,9}$ E      & 246600.0120 &  40.0   & 190.3465 & 0.039(6) & 0.586 & 0.502 \\ 
                 & $20_{10,11}-19_{10,10}$ A     & 246613.3920 &  80.0   & 190.3410 & 0.054(6)\tablenotemark{c} & 0.586 & 0.502 \\ 
                 & $20_{10,10}-19_{10,9}$ A      & 246613.3920 &          &          &          &       &       \\ 
                 & $20_{10,11}-19_{10,10}$ E     & 246623.1900 &  40.0   & 190.3340 & 0.032(6) & 0.586 & 0.502 \\ 
CH$_3$OCH$_3$    & $14_{1,14}-13_{0,13}$ EA      & 258548.8190 & 679   &  93.3326 & 0.082(7)\tablenotemark{c} & 0.561 & 0.480 \\
                 & $14_{1,14}-13_{0,13}$ AE      & 258548.8190 &          &          &          &       &       \\
                 & $14_{1,14}-13_{0,13}$ EE      & 258549.0630 &          &          &          &       &       \\
                 & $14_{1,14}-13_{0,13}$ AA      & 258549.3080 &          &          &          &       &       \\
                 & $18_{5,13}-18_{4,14)}$ AE      & 257911.0360 & 357   & 190.9744 & 0.058(8)\tablenotemark{c} & 0.560 & 0.481 \\ 
                 & $18_{5,13}-18_{4,14}$ EA      & 257911.1750 &          &          &          &       &       \\
                 & $18_{5,13}-18_{4,14}$ EE      & 257913.3120 &          &          &          &       &       \\
OCS              & $J = 19-18$                    & 231060.9934 &   9.72   &  110.8999 & 0.159(6) & 0.470 & 0.391 \\ 
CH$_2$DOH        & $8_{2,6}-8_{1,7}$ e$_0$       & 234471.0333 &   9.55   &  93.6659 & 0.043(6) & 0.467 & 0.387 \\ 
                 & $4_{1,4}-4_{1,3}$ e$_1$-e$_0$   & 246973.1071 &   1.10   &  37.6942 & 0.031(5) & 0.585 & 0.501 \\ 
                 & $3_{2,1}-3_{1,2}$ e$_0$          & 247625.7463 &   2.36   &  29.0149 & 0.027(7) & 0.584 & 0.500 \\ 
                 & $5_{2,4}-5_{1,5}$ e$_0$          & 261687.3662 &   4.01   &  48.3132 & 0.043(7) & 0.553 & 0.477 \\ 
NH$_2$CHO        & $12_{0,12}-11_{0,11}$        & 247390.7190 & 156   &  78.1227 & $<$0.007   & 0.585 & 0.501 \\
CH$_3$CHO        & $11_{1,10}-10_{1,9}$ E        & 216581.9304 &  69.0   &  64.8710 & $<$0.006   & 0.504 & 0.417 \\ 
C$_2$H$_5$OH     & $14_{3,11}-13_{3,10}$        & 246414.7897 &  21.6   & 155.7234 & $<$0.006   & 0.587 & 0.503 \\
C$_2$H$_5$CN     & $27_{2,25}-26_{2,24}$        & 246268.7367 & 398   & 169.8051 & $<$0.006   & 0.587 & 0.503 \\ 
CH$_3$COCH$_3$   & $23_{2,21}-22_{2,20}$        & 247562.2435 & 2428  & 153.3326 & $<$0.007\tablenotemark{c}   & 0.585 & 0.500 \\ 
                 & $23_{2,21}-22_{3,20}$        & 247562.2435 &  323  &          &          &       &       \\
                 & $23_{3,21}-22_{2,20}$        & 247562.2435 &  323  &          &          &       &       \\
                 & $23_{3,21}-22_{3,20}$        & 247562.2435 & 2428  &          &          &       &       \\
HCOOH            & $11_{2,10}-10_{2,9}$         & 246105.9739 &  21.5  &  83.7410 & $<$0.006   & 0.587 & 0.503 \\ 
CH$_2$(OH)CHO    & $23_{0,23}-22_{1,22}$        & 233037.3570 & 117  &  136.5843 & $<$0.006\tablenotemark{c}   & 0.470 & 0.388 \\
                 & $23_{1,23}-22_{0,22}$        & 233037.7300 & 117  &  136.5847 &          &       &       \\ 
\enddata
\tablenotetext{a}{Integrated intensity. Numbers in parentheses represent $1 \sigma$ which applies to the last significant digits. The upper limit also represents $1 \sigma$.}
\tablenotetext{b}{Synthesized beam size. The position angle is $112^{\circ}- 114^{\circ}$ for Setup 1 ($216-234$ GHz) and $93^{\circ}-95^{\circ}$ for Setup 2 ($245-260$ GHz).}
\tablenotetext{c}{Total integrated intensity over blended lines.}
\end{deluxetable}

\normalsize
\begin{table}
\caption{Column Densities of iCOMs \label{tab_CB68_COM_coldens}}
\begin{tabular}{lccc}
\hline \hline 
Molecule        &  \multicolumn{3}{c}{Column Density (cm$^{-2}$)} \\
                &  $T=100$ K                      &  $T=131$ K                     &  $T=150$ K                    \\ \hline 
HCOOCH$_3$      & $(2.2 \pm 0.4) \times 10^{17} $ & $(2.3 \pm 0.3) \times 10^{17}$ & $(2.4 \pm 0.3) \times 10^{17}$ \\
CH$_3$OCH$_3$   & $1.5 \times 10^{17} $           & $1.7 \times 10^{17} $          & $1.9 \times 10^{17} $         \\
CH$_2$DOH       & $1.2 \times 10^{17} $           & $1.5 \times 10^{17} $          & $1.7 \times 10^{17} $         \\
NH$_2$CHO\tablenotemark{a}       & $< 1.1 \times 10^{15} $         &  $< 1.3 \times 10^{15} $       &  $< 1.5 \times 10^{15} $      \\
C$_2$H$_5$OH\tablenotemark{a}    & $< 5.4 \times 10^{16} $         &  $< 6.0 \times 10^{16} $       &  $< 6.4 \times 10^{16} $      \\
C$_2$H$_5$CN\tablenotemark{a}    & $< 3.8 \times 10^{15} $         &  $< 4.2 \times 10^{15} $       &  $< 4.7 \times 10^{15} $      \\
(CH$_3$)$_2$CO\tablenotemark{a}  & $< 9.8 \times 10^{15} $         &  $< 1.1 \times 10^{16} $       &  $< 1.3 \times 10^{16} $      \\
HCOOH\tablenotemark{a}           & $< 6.4 \times 10^{15} $         &  $< 7.8 \times 10^{15} $       &  $< 8.7 \times 10^{15} $      \\
CH$_3$CHO\tablenotemark{a}       & $< 1.1 \times 10^{16} $         &  $< 1.6 \times 10^{16} $       &  $< 1.9 \times 10^{16} $      \\
CH$_2$(OH)CHO\tablenotemark{a}   & $< 7.9 \times 10^{15} $         &  $< 8.5 \times 10^{15} $       &  $< 8.9 \times 10^{15} $      \\
CH$_3$OH\tablenotemark{b}        & & $(2.7 \pm 1.0) \times 10^{18}$ & \\ 
\hline 
\end{tabular}
\tablenotetext{a}{Upper limit to the column densities are derived for non-detected molecules.}
\tablenotetext{b}{The rotation temperature ($T$) is derived to be $131 \pm 11$ K. 
				The beam filling factor ($f$) is derived to be $0.022 \pm 0.003$.} 
\end{table}

\begin{sidewaystable}
\centering

\tiny{ 
\caption{Abundance Ratios of iCOMs Relative to HCOOCH$_3$ 
		\label{tab_ratio-hcooch3}}
\begin{tabular}{lccccccccccccc}
\hline \hline 
Molecule        & CB68            & IRAS16293\tablenotemark{\tiny a}  & IRAS16293\tablenotemark{\tiny b} & NGC1333\tablenotemark{\tiny c} & NGC1333\tablenotemark{\tiny d} & B335\tablenotemark{\tiny e} & BHR71\tablenotemark{\tiny f}  & HH212\tablenotemark{\tiny g}  & B1-c\tablenotemark{\tiny h} & HOPS-108\tablenotemark{\tiny i} & Ser-emb 8\tablenotemark{\tiny h}  & Ser-emb 11\tablenotemark{\tiny j}  &  SVS13A\tablenotemark{\tiny k}  \\
                &                 & Source A & Source B       & IRAS 2A  & IRAS 4A2 &       & IRS1   &    &    &    &    &    &    \\ \hline 
CH$_3$OCH$_3$   & $0.74 \pm 0.11$ & $1.9^{+1.0}_{-0.7}$ & $0.9^{+0.3}_{-0.2}$  & $0.6^{+0.4}_{-0.3}$  & $1.3^{+0.19}_{-0.09}$   & $0.73^{+0.12}_{-0.11}$ & $0.12^{+0.04}_{-0.03}$ & $-$ & $1.26^{+0.09}_{-0.08}$ & $0.20 \pm 0.10$ & $0.75^{+0.12}_{-0.10}$ & $0.69^{+0.08}_{-0.05}$ &  $0.33^{+0.08}_{-0.07}$  \\
CH$_3$CHO       & $< 0.07$ & $0.013^{+0.007}_{-0.005}$   & $0.46^{+0.15}_{-0.12}$         &   $-$   & $0.43^{+0.15}_{-0.14}$  & $0.54^{+0.10}_{-0.09}$ & $-$ &$0.5^{+0.5}_{-0.2}$ & $0.24 \pm 0.05$ &  $0.020 \pm 0.004$ & $0.063^{0.011}_{0.009}$  & $-$ & $0.08 \pm 0.05$ \\
NH$_2$CHO       & $< 0.006$  & $0.007^{+0.004}_{-0.003}$  &    $-$     & $0.15^{+0.05}_{-0.06}$  & $-$ & $0.092^{+0.014}_{-0.012}$ & $-$ & $0.05^{+0.06}_{-0.03}$ & $0.027 \pm 0.013$ & $0.003 \pm 0.001$ &  $0.017 \pm 0.003$  & $0.0050^{+0.0013}_{-0.0011}$ & $0.006 \pm 0.002$   \\
C$_2$H$_5$OH    & $< 0.3$   & $0.30^{+0.15}_{-0.11}$   & $0.9^{+0.3}_{-0.2}$         & $1.0 \pm 0.6$   &  $-$    & $0.81^{+0.16}_{-0.14}$ &  $0.16^{+0.05}_{-0.04}$ & $0.8^{+0.9}_{-0.5}$ & $0.79 \pm 0.16$  &  $-$ & $0.19^{+0.03}_{-0.02}$  & $0.145^{+0.018}_{-0.013}$ & $0.29^{+0.07}_{-0.11}$  \\
C$_2$H$_5$CN    & $< 0.02$    & $-$   &  $-$   & $0.019^{+0.006}_{-0.008}$ & $-$ & $0.037^{+0.008}_{-0.007}$ & $-$ & $-$ & $0.057 \pm 0.016$ & $0.006^{+0.004}_{-0.002}$ & $0.07 \pm 0.04$ & $0.0104^{+0.0019}_{0.0015}$ & $0.008^{+0.002}_{-0.003}$   \\
(CH$_3$)$_2$CO  & $< 0.05$  & $0.09^{+0.05}_{0.03}$   & $0.9^{+0.3}_{-0.2}$         &   $-$    & $0.15^{+0.0.03}_{-0.0.02}$  & $0.18 \pm 0.03$ & $0.12^{+0.04}_{-0.03}$ & $-$ & $-$ & $0.055 \pm 0.013$ & $0.35^{+0.05}_{-0.04}$  & $0.091^{+0.015}_{-0.012}$ & $0.19^{+0.12}_{-0.14}$  \\
HCOOH           & $< 0.03$  & $0.05^{+0.03}_{-0.02}$     & $0.22^{+0.07}_{-0.06}$         &   $-$   &  $-$    & $1.0 \pm 0.2$  & $-$ &  $-$ & $0.037 \pm 0.011$ & $-$ & $ 0.57^{+0.08}_{-0.06}$ & $-$ & $0.05 \pm 0.02$ \\ 
$L_{\rm bol}$/$L_\odot$ & 0.86\tablenotemark{\tiny l}  & 21\tablenotemark{\tiny m}  & 21\tablenotemark{\tiny m}  & 35.7\tablenotemark{\tiny n}  & 9.1\tablenotemark{\tiny n}  & 0.57\tablenotemark{\tiny o}  & 13.5\tablenotemark{\tiny p}  & 9\tablenotemark{\tiny g}  & 3.2\tablenotemark{\tiny q}  &       38.3\tablenotemark{\tiny r} & 5.4\tablenotemark{\tiny s}  & 4.8\tablenotemark{\tiny s}  & 32.5\tablenotemark{\tiny t}  \\ 
$T_{\rm bol}$/K & 41\tablenotemark{\tiny l}  & 43\tablenotemark{\tiny u}  & 43\tablenotemark{\tiny u}  & 50\tablenotemark{\tiny n}  & 33\tablenotemark{\tiny n}  & 36\tablenotemark{\tiny n}  & 51\tablenotemark{\tiny p} & $\sim 50$\tablenotemark{\tiny v} & 46\tablenotemark{\tiny q} & 39\tablenotemark{\tiny r} & 58\tablenotemark{\tiny s}  & 77\tablenotemark{\tiny s}  & 188\tablenotemark{\tiny t}   \\ 
SED Class & Class 0 & Class 0 & Class 0 & Class 0 & Class 0 & Class 0 & Class 0 & Class 0 & Class 0 & Class 0 & Class 0 & Class I & Class I \\
\hline 
\end{tabular}
\tablenotetext{}{$^a$\citet{man20}. $^b$\citet{jor18}. $^c$\citet{taq15}. $^d$\citet{Lop17}. $^e$\citet{ima16}. $^f$\citet{yan20}. $^g$\citet{lee19}. $^h$\citet{van20} for O-bearing COMs and \citet{naz21} for N-bearing COMs. $^i$\citet{cha22}. $^j$\citet{mar21}. $^k$\citet{yan21}. $^l$\citet{lau13}. $^m$\citet{jor16}. Source A and Source B are not resolved. $^n$\citet{kri12}. A1 and A2 are not resolved. $^o$\citet{eva15}. $^p$\citet{yan18}. $^q$\citet{kar18}. $^r$\citet{tob19}. $^s$\citet{eno11}. $^t$\citet{tob16}. $^u$\citet{jor02}. Source A and Source B are not resolved. $^v$\citet{lee06}. Estimated from $T_{\rm{dust}}$.}
}

\end{sidewaystable}

\begin{table}
\centering

\caption{Comparison with Chemical Model$^a$
		\label{tab_iCOMs_Garrod}}
\begin{tabular}{lcccccc}
\hline \hline 
Abundance Ratio             & CB68   & Fast\tablenotemark{a}  & Medium\tablenotemark{a}  & Slow\tablenotemark{a}  & Fiducial\tablenotemark{b} & $T_{\rm min} = 20$ K\tablenotemark{b} \\ \hline 
CH$_3$OCH$_3$/HCOOCH$_3$ & 0.74 $\pm$ 0.11 &  0.58     & 0.90     & 0.89 & 62 & 1.7 \\
CH$_3$CHO/HCOOCH$_3$     & $<$0.07         &  0.21    & 0.38    & 1.0  & 8.4 & 61 \\
NH$_2$CHO/HCOOCH$_3$     & $<$0.006        &  0.10     & 0.13    & 0.21  & 42 & 290 \\
\hline 
\end{tabular}
\tablenotetext{a}{Taken from \citet{gar13}. 
			Results for the fast, medium, and slow warm-up cases are shown.} 
\tablenotetext{b}{Taken from \citet{aik20}. 
			Results for the fiducial model and the model of the minimum temperature during the static phase of 20 K are listed.}
\end{table}

\end{document}